\documentclass{article}

\usepackage{arxiv}

\usepackage[utf8]{inputenc} 
\usepackage[T1]{fontenc}    
\usepackage{hyperref}       
\usepackage{url}            
\usepackage{booktabs}       
\usepackage{amsfonts}       
\usepackage{nicefrac}       
\usepackage{microtype}      
\usepackage{lipsum}
\usepackage{graphicx}
\usepackage{amssymb}
\usepackage{latexsym}
\usepackage{amsmath}
\usepackage{caption}
\usepackage[affil-it]{authblk}
\usepackage[numbers]{natbib}
\usepackage{multirow}
\DeclareCaptionLabelFormat{Sfigure}{\textbf{Figure S\thefigure}}

\title{Attention incorporated network for sharing low-rank, image and k-space information during MR image reconstruction to achieve single breath-hold cardiac Cine imaging}

\author[1]{Siying Xu\thanks{Corresponding author. Email: siying.xu@med.uni-tuebingen.de}}
\author[2]{Kerstin Hammernik}
\author[3]{Andreas Lingg}
\author[3]{Jens Kübler}
\author[3]{Patrick Krumm}
\author[2,4,5]{Daniel Rueckert}
\author[1,6]{Sergios Gatidis}
\author[1]{Thomas Küstner}

\affil[1]{Medical Image and Data Analysis (MIDAS.lab), Department of Diagnostic and Interventional Radiology, University of Tuebingen, Tuebingen, Germany}
\affil[2]{School of Computation, Information and Technology, Technical University of Munich, Munich, Germany}
\affil[3]{Department of Diagnostic and Interventional Radiology, University of Tuebingen, Tuebingen, Germany}
\affil[4]{Klinikum Rechts der Isar, Technical University of Munich, Munich, Germany}
\affil[5]{Department of Computing, Imperial College London, London, United Kingdom}
\affil[6]{Department of Radiology, Stanford University, Stanford, California, USA}

\begin{document}
\maketitle

\begin{abstract}
Cardiac Cine Magnetic Resonance Imaging (MRI) provides an accurate assessment of heart morphology and function in clinical practice. However, MRI requires long acquisition times, with recent deep learning-based methods showing great promise to accelerate imaging and enhance reconstruction quality. Existing networks exhibit some common limitations that constrain further acceleration possibilities, including single-domain learning, reliance on a single regularization term, and equal feature contribution. To address these limitations, we propose to embed information from multiple domains, including low-rank, image, and k-space, in a novel deep learning network for MRI reconstruction, which we denote as A-LIKNet. A-LIKNet adopts a parallel-branch structure, enabling independent learning in the k-space and image domain. Coupled information sharing layers realize the information exchange between domains. Furthermore, we introduce attention mechanisms into the network to assign greater weights to more critical coils or important temporal frames. Training and testing were conducted on an in-house dataset, including 91 cardiovascular patients and 38 healthy subjects scanned with 2D cardiac Cine using retrospective undersampling. Additionally, we evaluated A-LIKNet on the real-time $8\times$ prospectively undersampled data from the OCMR dataset. The results demonstrate that our proposed A-LIKNet outperforms existing methods and provides high-quality reconstructions. The network can effectively reconstruct highly retrospectively undersampled dynamic MR images up to $24\times$ accelerations, indicating its potential for single breath-hold imaging.\footnote{Code available: \href{https://github.com/midas-tum/A-LIKNet}{https://github.com/midas-tum/A-LIKNet}}
\end{abstract}


\section{Introduction}
\label{sec1}
For cardiovascular disease identification, an accurate analysis of cardiac function and anatomy is required. Cine magnetic resonance imaging (MRI) is commonly used in clinical practice as a non-invasive and radiation-free assessment of the structural dynamics over the cardiac cycle. However, conventional multi-slice 2D cardiac Cine is acquired under multiple breath-holds, which not only results in a longer acquisition time but also causes patient discomfort and respiratory-induced slice misalignments. Therefore, numerous efforts have been made to accelerate cardiac Cine MRI over the past decades.

Parallel imaging (PI)~\citep{pruessmann1999sense, griswold2002generalized, lustig2010spirit, uecker2014espirit} utilizes the recorded regular undersampled k-space signals over spatially distributed MR receiver coils. By using spatial information over the coils, the image can be reconstructed. PI reconstruction can be either performed in the image domain, such as SENSitivity Encoding (SENSE)~\citep{pruessmann1999sense}, or in the k-space, such as GeneRalized Auto-calibrating Partially Parallel Acquisitions (GRAPPA)~\citep{griswold2002generalized}. In practice, the acceleration is often set to a lower value than theoretically possible, which is limited by the number of receiver coils, to ensure satisfactory image quality. For cardiac Cine in clinical routine, $2\times$ to $4\times$ acceleration rates are commonly employed using parallel imaging techniques. Although PI enables the reduction of scan time, its acceleration capability is limited due to the fact that excessive accelerations can lead to enhanced noise.

Compressed sensing (CS) has been proposed as an alternative to accelerate MR imaging, which exploits the fact that MR images can be sparsely represented in an appropriate transform domain. Given incoherent aliasing artifacts in the transform domain due to a random k-space undersampling, the image can be reconstructed by non-linear optimization. Representative sparsity-enforcing methods include Wavelet transform~\citep{chaari2011wavelet}, Total Variation (TV)~\citep{osher2005iterative, block2007undersampled}, and dictionary learning~\citep{liu2013adaptive, caballero2014dictionary, weller2016reconstruction}. CS MRI can also be combined with PI to jointly use image sparsity and coil sensitivity information~\citep{lustig2007sparse, liu2008sparsesense}, thus further increasing the imaging speed. However, CS MRI still faces several challenges, such as long reconstruction time due to the iterative reconstruction process, challenging selection of hyperparameters for the sparse regularizations, and high computational burden for high-dimensional MRI reconstruction.

Similar to the goal of CS, low-rank priors aim to represent data or signals in a compact form by reducing the rank of a matrix. Methods that jointly consider sparse and low-rank priors have shown improved performance in dynamic MR image reconstruction. These methods can be categorized into two main approaches: low-rank and sparse (L\&S)~\citep{lingala2011accelerated, zhao2012image} and Low-rank plus sparse (L+S)~\citep{candes2011robust, chandrasekaran2011rank, otazo2015low}. L\&S assumes that MR images simultaneously exhibit low-rank and sparse characteristics in the specific domain, while L+S hypothesizes that the image can be decomposed into a superposition of a low-rank component and a sparse component~\citep{otazo2015low}. In dynamic MR imaging, L typically represents the slowly varying background, while S models the dynamic information superimposed on top of it. Although the joint consideration of sparse and low-rank priors has further improved the reconstruction speed and image quality, these methods often require iterative optimization algorithms to estimate the low-rank and sparse components, leading to high computational complexity and longer reconstruction time. Additionally, determining the appropriate parameters relies on empirical observations. The performance of these methods is heavily dependent on parameter selection, and different applications necessitate different parameter settings.

Recently, deep learning (DL) reconstruction methods started to flourish with the developments of graphics processing units (GPU) and the increasing availability of large-scale databases. Based on the operation domain and the inputs to the neural network, supervised DL-based MRI reconstruction methods can be classified into five categories: image enhancement~\citep{lee2018deep, hauptmann2019real, kofler2019spatio}, k-space learning~\citep{cheng2018deepspirit, akccakaya2019scan, han2019k}, physics-based reconstruction~\citep{yang2016deep, schlemper2017deep, hammernik2018learning, aggarwal2018modl, qin2018convolutional, duan2019vs, kustner2020cinenet}, plug and play priors (PnP)~\citep{venkatakrishnan2013plug, wang2016accelerating, meinhardt2017learning, yazdanpanah2019deep, ahmad2020plug, liu2020rare}, and distribution learning~\citep{jalal2021instance, chung2022score, levac2023conditional}. 

Image enhancement networks take noise-corrupted images as inputs and artifact-free images as labels in a supervised setting. In contrast, k-space learning, such as RAKI~\citep{akccakaya2019scan}, operates in k-space, utilizing acquired k-space data to estimate nonlinear mapping kernels between coils to fill in missing k-space data - following the concept of PI. However, these methods do not ensure data consistency in the acquired k-space samples. Hence, physics-based unrolled networks have been proposed, alternating between the neural network and a data consistency layer in an iterative optimization process. The neural network automatically learns the regularization in the image domain, while the intermittent data consistency layers introduce similarity to the acquired k-space data. Typical methods such as deep cascade network~\citep{schlemper2017deep}, MoDL~\citep{aggarwal2018modl}, variational network (VN)~\citep{hammernik2018learning}, and CINENet~\citep{kustner2020cinenet} greatly improved the reconstruction speed and image quality in highly undersampled data. However, the reconstruction performance can be limited when only operating in one domain, either in the image domain or k-space. Hybrid physics-based reconstruction approaches~\citep{singh2022joint, jun2021joint} such as KIKI-net~\citep{eo2018kiki}, MD-CNN~\citep{el2021multi}, and KV-Net~\citep{liu2022dual} additionally introduce k-space subnetworks that integrate learning in both k-space and image domains. PnP algorithms embed prior information, typically represented by a denoiser, into a larger optimization algorithm to avoid dependence on the forward physics model during training. Distribution learning utilizes generative models such as diffusion models to learn more accurate prior knowledge about data structures and data distributions for reconstruction. Furthermore, learning-based low-rank methods~\citep{huang2021deep, ke2021learned, wang2022one} allow for learning hyperparameters in traditional low-rank methods, enhancing flexibility and reliability. 

While DL-based reconstruction methods have shown significant improvements compared to PI and CS MRI, we observe that existing learning-based methods have at least one of the following challenges: (1) \textit{Operation on a single domain:} image enhancement, k-space learning, PnP and low-rank methods focus on single-domain reconstruction, meaning that the neural network operates either in the k-space or image domain. Despite the utilization of k-space samples in the data consistency layers in physics-based unrolled networks, they do not explicitly learn k-space information. (2) \textit{Focus on single prior:} the image network in physics-based unrolled networks is solely for learning the sparse prior, disregarding the inherent low-rank characteristic of dynamic images that could be exploited. (3) \textit{Equal feature contribution:} in most existing DL-based reconstruction methods, the features within the neural network are treated equally in all dimensions, which impairs the representation ability of the network. In some works, the attention mechanism has been explored in MR image reconstruction~\citep{huang2019mri, li2021modified, li2021high}, but they mainly focus on spatial or network-channel-wise feature attention. The importance of temporal features for dynamic images and coil-wise features for k-space data has not been explored.

To address the aforementioned limitations and improve the reconstruction of dynamic MR imaging, we propose A-LIKNet, which incorporates the \textbf{A}ttention mechanism to share \textbf{L}ow-rank, \textbf{I}mage, and \textbf{K}-space information during reconstruction. Unlike most existing networks with a single-input, single-output structure, our A-LIKNet consists of parallel k-space and image branches that output the reconstructed k-space and image. Both branches contribute to the loss calculation, allowing for the simultaneous reconstruction of k-space and image. The parallel structure ensures the independence of single-domain learning while simultaneously enabling attention-based sharing between domains. Therefore, we introduce a learnable information sharing layer (ISL) in each optimization iteration to maximize multi-domain information sharing. In the image branch, we consider both the sparsity and low-rank nature of dynamic MR images. To achieve this, we employ a learnable image sub-network and low-rank sub-network for the sparse and low-rank priors, respectively. Furthermore, to enhance the representational capacity of the neural network, we introduce attention mechanisms along the temporal dimension in the image sub-network and the coil dimension in the k-space sub-network.

The main contributions of this work can be summarized as follows: (1) \textit{Maximizing information utilization in multiple domains:} the novel parallel-branch architecture with intermittent ISLs maximizes the information sharing between domains while keeping the independence of image and k-space learning. The k-space branch complements global coil-resolved learning to the local receptive field confinement of the image sub-network. (2) \textit{Leveraging the low-rank property of dynamic images:} we design a local spatial-temporal low-rank threshold learning in the low-rank sub-network. (3) \textit{Application of attention mechanism:} we introduce a time-wise attention block in the image sub-network and a coil-wise attention block in the k-space sub-network, allowing the network to assign greater importance to specific frames or coils. Investigations and comparisons to related DL reconstructions~\citep{lustig2007sparse, lingala2011accelerated, aggarwal2018modl, eo2018kiki, huang2021deep} were carried out on retrospectively undersampled data in an in-house database of 91 patients and 38 healthy subjects, as well as evaluations of pre-trained models on the prospectively undersampled OCMR dataset~\citep{chen2020ocmr}. Experiments demonstrate that the proposed A-LIKNet can reconstruct undersampled cardiac Cine images under extremely high acceleration factors up to $24\times$, showcasing the potential for single breath-hold imaging.

\section{Methodology}
In this section, we will first introduce the mathematical background of the proposed A-LIKNet in Sec.~\ref{sec: problem_formulation}, explaining how we formulate and solve the MR image reconstruction problem. In Sec.~\ref{sec: proposed network}, we will describe how we transform the mathematical expressions into the various components of our proposed A-LIKNet.

\subsection{Problem formulation}\label{sec: problem_formulation}
Let $\mathbf{x}\in\mathbb{C}^n$ be the desired fully-sampled complex-valued dynamic MR image stacked as a vector, with $n$ being the number of pixels in $\mathbf{x}$. Let $\mathbf{y}_{u}\in\mathbb{C}^m$ denote the undersampled measurements in k-space, with $m$ being the number of sampled k-space points and $m < n$. The forward model links $\mathbf{x}$ and $\mathbf{y}_{u}$:
\begin{equation}
\label{eq:imaging}
    \mathbf{y}_{u}=\mathbf{Ax},
\end{equation}
where the encoding operator $\mathbf{A}\in\mathbb{C}^{m\times n}$ is describing the MR imaging operations:
\begin{equation}\label{eq:ForwardOp}
    \mathbf{A}=\mathbf{MFS}.
\end{equation}
Here, $\mathbf{S}$ refers to the coil sensitivity maps of the multi-coil imaging scenario, $\mathbf{F}$ is the Fourier transform operator, and $\mathbf{M}$ denotes the binary undersampling trajectory. As the sampling process violates the Nyquist-Shannon theorem, the reconstruction problem in which we try to recover the image $\mathbf{x}$ from the acquired k-space $\mathbf{y}_{u}$ is ill-posed. To solve this problem, we transform the reconstruction into a regularized reconstruction problem:
\begin{equation}\label{eq:cs_solution}
    \mathbf{\hat{x}}=\arg\ \underset{\mathbf{x}}{\min}\ \frac{1}{2}\parallel\mathbf{Ax}-\mathbf{y}_{u}\parallel_{2}^{2}+\lambda R\left(\mathbf{x}\right),
\end{equation}
where $\lambda$ weights the contribution of the regularization term $R(\cdot)$. In general, Total Variation and $l_{1}$-norm on the image $\mathbf{x}$ are used as the sparse regularization $R(\cdot)$. In dynamic MRI, low-rankness is an additional important prior that can be learned and leveraged during reconstruction. Therefore, we take into account both the sparsity of images and the inherent low-rank property of dynamic images, extending Eq.~\ref{eq:cs_solution} to:
\begin{equation}\label{eq:img_solution}
    \mathbf{\hat{x}}=\arg\,\underset{\mathbf{x}}{\min}\ \frac{1}{2}\parallel\mathbf{Ax}-\mathbf{y}_{u}\parallel_{2}^{2}+\lambda_{s}R_{s}\left(\mathbf{x}\right)+\lambda_{l}R_{l}\left(\mathbf{x}\right),
\end{equation}
where $R_{s}(\cdot)$ is the sparse regularization term with weighting factor $\lambda_{s}$, and $R_{l}(\cdot)$ is the low-rank regularizer with weighting parameter $\lambda_{l}$.

In our proposed method, we not only focus on the reconstruction of the coil-combined image but also simultaneously aim at reconstructing the coil-resolved k-space data. Therefore, in addition to Eq.~\ref{eq:img_solution}, we introduce the minimization problem in k-space:
\begin{equation}\label{eq:kspace_solution}
    \mathbf{\hat{y}}=\arg\,\underset{\mathbf{y}}{\min}\ \frac{1}{2}\parallel\mathbf{My}-\mathbf{y}_{u}\parallel_{2}^{2}+\lambda_{k} R_{k}\left(\mathbf{y}\right),
\end{equation}
where $R_{k}(\cdot)$ represents the frequency domain regularization term weighted by $\lambda_{k}$.

In the ideal scenario where we can obtain the globally optimal solution, Eq.~\ref{eq:img_solution} and Eq.~\ref{eq:kspace_solution} would be equivalent and result in the same solution. However, when we separately solve Eq.~\ref{eq:img_solution} and Eq.~\ref{eq:kspace_solution}, we often end up at local minima for which $\mathbf{\hat{x}}$ and $\mathbf{\hat{y}}$ do not correspond. To avoid this situation and maximize information sharing between domains, we constrain the reconstructed image and k-space by the forward model:
\begin{equation}\label{eq:ISL_constrain}
    \mathbf{\hat{y}}=\mathbf{FS\hat{x}}.
\end{equation}
Eq.~\ref{eq:ISL_constrain} implies that we aim for a solution where $\mathbf{\hat{x}}$ and $\mathbf{\hat{y}}$ correspond to each other.

\begin{figure}[!t]
\includegraphics[width=\linewidth]{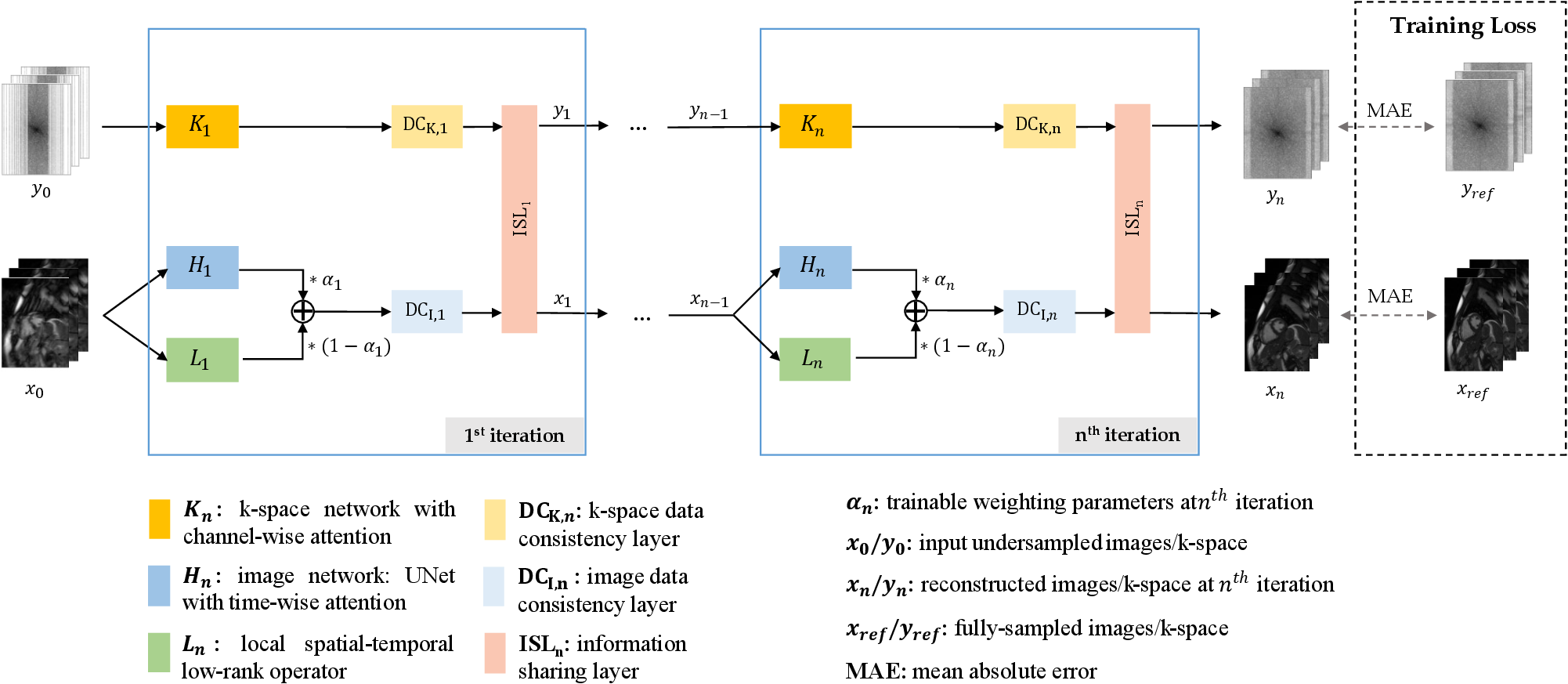}
\caption{Proposed A-LIKNet: Attention-incorporated physics-based unrolled network for sharing low-rank, image, and k-space information during MR image reconstruction. The network consists of two branches: the bottom branch, featuring an image subnetwork, a learnable local low-rank operator, and an image data consistency layer for solving Eq.~\ref{eq:img_solution}, and the upper branch, comprising a k-space subnetwork and a k-space data consistency layer to address Eq.~\ref{eq:kspace_solution}. The information sharing layer facilitates communication between branches, enforcing consistency as in Eq.~\ref{eq:ISL_constrain}.}
\label{fig:network}
\end{figure}

In summary, Eq.~\ref{eq:img_solution}, \ref{eq:kspace_solution}, and \ref{eq:ISL_constrain} form the overall reconstruction problem. Eq.~\ref{eq:img_solution} corresponds to the image branch in Fig.~\ref{fig:network}, which is designed to seek the optimal solution in the image domain. Eq.~\ref{eq:kspace_solution} is solved by the k-space branch in Fig.~\ref{fig:network}, which is responsible for recovering the k-space data in the frequency domain. As each data point in k-space contains the corresponding frequency and phase information of all pixels in the image domain, the inclusion of the k-space branch not only leverages coil-resolved information but also compensates for the limited receptive field of view in the image domain. Moreover, Eq.~\ref{eq:ISL_constrain} is implemented by the information sharing layer in Fig.~\ref{fig:network}. By harnessing estimated coil sensitivity information, ISL facilitates the exchange of information across domains. We hypothesize that when a particular branch becomes trapped in a local minimum or saddle point, this information sharing aids in guiding the model toward the global minimum. In the following sections, we will provide detailed derivations and descriptions of each network component in A-LIKNet.

\subsection{Components of A-LIKNet}\label{sec: proposed network}
The overall architecture of A-LIKNet is shown in Fig.~\ref{fig:network}, which is built as a physics-based unrolled reconstruction network and consists of an image branch (Sec.~\ref{sec: image branch}), a k-space branch (Sec.~\ref{sec: ksp branch}), and information sharing layers (Sec~\ref{sec: ISL}) between these two branches.

\subsubsection{Image branch}\label{sec: image branch}
The image branch is constructed to solve Eq.~\ref{eq:img_solution} and consists of an image subnetwork, a low-rank subnetwork, and a data consistency layer. By introducing auxiliary variables $\mathbf{p}\in\mathbb{C}^n$ and $\mathbf{q}\in\mathbb{C}^n$, Eq.~\ref{eq:img_solution} can be equivalently written as:
\begin{equation}
    \begin{array}{c}
         \mathbf{\hat{x}} = \arg\,\underset{\mathbf{x}}{\min}\ \frac{1}{2}\parallel\mathbf{Ax}-\mathbf{y}_{u}\parallel_{2}^{2}+\lambda_{s}R_{s}\left(\mathbf{p}\right)+\lambda_{l}R_{l}\left(\mathbf{q}\right) \\ \mathrm{s.t.} \ \mathbf{p} = \mathbf{x}, \mathbf{q} = \mathbf{x},  
    \end{array}
\end{equation}
which can be transformed into the unconstrained Lagrangian function using half quadratic splitting (HQS) with auxiliary variables $\mu_{1}$ and $\mu_{2}$:
\begin{equation}\label{eq:lang_form}
    \begin{split}
        \textit{L}_{\mu_{1},\mu_{2}}(\mathbf{x},\mathbf{p},\mathbf{q})&=\frac{1}{2}\parallel\mathbf{Ax}-\mathbf{y}_{u}\parallel_{2}^{2}+\lambda_{s}R_{s}\left(\mathbf{p}\right)+\lambda_{l}R_{l}\left(\mathbf{q}\right) \\ 
        &+\frac{\mu_{1}}{2}\left\|\mathbf{p}-\mathbf{x}\right\|^{2}_2+\frac{\mu_{2}}{2}\left\|\mathbf{q}-\mathbf{x}\right\|^{2}_2.
    \end{split}
\end{equation}
This problem can be solved iteratively to obtain the following sub-problems:
\begin{equation}\label{eq:iter_sol}
    \left\{
        \begin{aligned}
        \mathbf{p}_{n+1} = {\arg}\,\underset{\mathbf{p}}{\min}\ &\frac{\mu_{1}}{2}\left\|\mathbf{p}-\mathbf{x}_{n}\right\|^{2}_2+\lambda_{s}R_{s}\left(\mathbf{p}\right)\\
        \mathbf{q}_{n+1} = {\arg}\,\underset{\mathbf{q}}{\min}\ &\frac{\mu_{2}}{2}\left\|\mathbf{q}-\mathbf{x}_{n}\right\|^{2}_2+\lambda_{l}R_{l}\left(\mathbf{q}\right)\\
        \mathbf{x}_{n+1} = {\arg}\,\underset{\mathbf{x}}{\min}\ &\frac{1}{2}\left||\mathbf{Ax}-\mathbf{y}_{u}\right\|_{2}^{2}+\frac{\mu_{1}}{2}\left\|\mathbf{p}_{n+1}-\mathbf{x}\right\|^{2}_2 \\
        &+\frac{\mu_{2}}{2}\left\|\mathbf{q}_{n+1}-\mathbf{x}\right\|^{2}_2.\\
        \end{aligned}
    \right.
\end{equation}

To solve $\mathbf{p}_{n}$, we introduce an image subnetwork $\mathbf{H}_{n}$, a convolutional neural network (CNN) designed to learn a sparse prior for the given current image $\mathbf{x}_{n-1}$. $\mathbf{H}_{n}$ intends to reconstruct local information on a coarse-to-fine scale. To adaptively learn a low-rank prior to solve $\mathbf{q}_{n}$, we design a low-rank subnetwork, denoted as $\mathbf{L}_{n}$. The subnetwork $\mathbf{L}_{n}$ decomposes a given $\mathbf{x}_{n-1}$ into local spatial-temporal patches and learns the singular value threshold for each patch to constrain the low-rank nature. Following the updates of $\mathbf{p}$ and $\mathbf{q}$, the data consistency operator $\mathbf{DC}_{\mathrm{I} ,n}$ will be employed to compute the output $\mathbf{x}_{n}$ for the current iteration, ensuring data fidelity. The subscript $\mathrm{I}$ indicates the DC layer operation in the image domain. With the learnable image subnetwork $\mathbf{H}_{n}$, the low-rank subnetwork $\mathbf{L}_{n}$, and the data consistency layer $\mathbf{DC}_{\mathrm{I} ,n}$, the equations in Eq.~\ref{eq:iter_sol} can be reformulated as follows:
\begin{equation}\label{eq:iter_sol_model}
    \left\{
    \begin{array}{l}
    \mathbf{p}_{n}=\mathbf{H}_{n}(\mathbf{x}_{n-1})\\\mathbf{q}_{n}=\mathbf{L}_{n}(\mathbf{x}_{n-1})\\\mathbf{x}_{n}=\mathbf{DC}_{\mathrm{I} ,n}(\mathbf{p}_{n}, \mathbf{q}_{n}),
    \end{array}\right.
\end{equation}
where $n\in \left \{ 1,...,N \right \} $, with $N$ being the total iteration number or unrolls of the network modules.

\begin{figure}[!t]
\includegraphics[width=\linewidth]{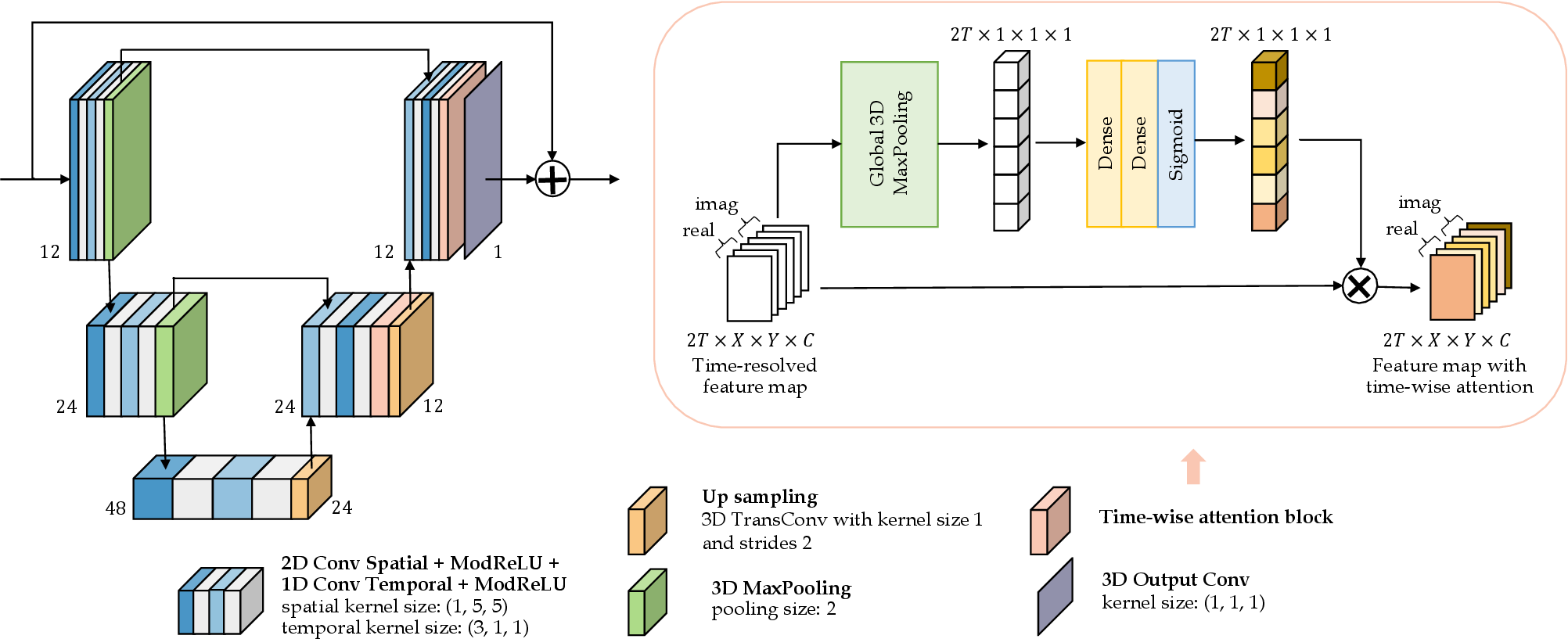}
\caption{Image subnetwork with time-wise attention block. A residual 2D+t UNet with attention blocks in the decoder. The number of filters is denoted beside each layer. The attention block squeezes and excites spatial-channel features to generate a temporal attention map, assigning different weights to frames.}
\label{fig:image_network}
\end{figure}

\paragraph{Image subnetwork}
In the first step of Eq.~\ref{eq:iter_sol_model}, a complex-valued residual 2D+t UNet $\mathbf{H}_{n}$ with time-wise attention blocks is used to learn the sparse prior. As depicted in Fig.~\ref{fig:image_network}, the UNet contains two stages. In each stage, a 2D+t convolution, i.e., a 2D spatial convolution followed by a 1D temporal convolution, is performed. A ModReLU activation function~\citep{arjovsky2016unitary} is introduced between and after convolutions. Using a 2D+t convolution instead of 3D convolution not only reduces the computational burden but also allows for independent learning in the spatial and temporal domains. At the end of each encoder stage, a 3D max pooling layer is used to reduce the feature size. The upsampling operation in the decoder is implemented by a 3D transpose convolution. The final output layer has only one kernel with a linear activation function, serving to predict the coil-combined image. Residual paths between the encoder/decoder stages and the input/output enable the network to focus solely on learning to remove the noise and aliasing artifacts from the images, reducing the learning complexity.

Different from the conventional UNet\citep{ronneberger2015unet}, we incorporate the attention mechanism along the temporal dimension. As shown in Fig.~\ref{fig:image_network}, the time-wise attention block is added at the end of each decoder stage. This block leverages the squeeze-and-excitation mechanism~\citep{hu2018squeeze}. Specifically, we concatenate the real and imaginary parts of the time-resolved feature maps along the temporal dimension to form real-valued arrays. Subsequently, 3D global max pooling is performed over the spatial and channel dimensions to squeeze the feature maps. The squeezed spatial-channel information is then processed through two fully connected layers followed by a sigmoid function to generate an attention map. This map excites the original features by multiplication, assigning distinct weights to different temporal features. The newly enhanced feature map is transformed back into the complex domain before being passed to the subsequent layers of the UNet.

Due to MR images being complex-valued, most existing methods tend to convert them into magnitude or two-channel images (i.e., not considering the relationship between real and imaginary components). In contrast, except for the attention block, our network is built upon complex-valued operations, including complex convolution~\citep{trabelsi2018deep} and complex activation functions~\citep{arjovsky2016unitary}. The complex operations preserve both the phase and magnitude information of the input signal. Furthermore, jointly processing the real and imaginary components within one operation conserves computational resources and mitigates information loss.

\paragraph{Low-rank subnetwork}
The low-rank subnetwork is employed to update $\mathbf{q}$ in the second step of Eq.~\ref{eq:iter_sol_model}. In dynamic MR images, each pixel strongly correlates to the same or adjacent pixels in the neighboring time frames. Thus, our low-rank subnetwork is tailored to focus on local spatial-temporal relationships. We split the input image sequence $\mathbf{x}_{n-1}$ into spatial-temporal patches. Specifically, the $T$ frames are divided into $n_{t}$ groups, and the spatial domain is segmented into $n_{x}\times n_{y}$ patches. It is important to note that due to the varying spatial sizes of input images, we define the number of partitions rather than specific patch sizes. Here, $n_x$, $n_y$, and $n_t$ refer to the number of patches in each dimension. To ensure information continuity and prevent disjoint edge effects, these patches overlap in the spatial domain, and the size of the overlapping regions is adaptively calculated based on the image size. Decomposing the image into patches brings the additional advantage of increasing computational efficiency, especially in the case of long dynamic sequences.

Low-rank regularization is applied to each patch. The implementation is based on the low-rank layer in L+S-Net~\citep{huang2021deep}, with the key distinction in threshold calculation and patch-wise application. For the $i^{th}$ patch $\mathbf{x}_{i}$, we perform singular value decomposition (SVD):
\begin{equation}
    \mathbf{x}_{i} = \mathbf{U\Sigma V }^{H},  
\end{equation}
where the singular value matrix $\mathbf{\Sigma } =\mathrm{diag}(\sigma _{j} ), 1\le j\le r$ with $r=\operatorname{rank}(\mathbf{x}_{i} )$ contains singular values of $\mathbf{x}_{i}$ along the diagonal, and $\mathbf{U}, \mathbf{V}$ contain the singular vectors. We assign each patch a learnable singular value threshold coefficient $\tau_{i}$. By applying a sigmoid function to $\tau_{i}$ and multiplying the result with the maximum singular value, we ensure that the learned threshold lies between 0 and the existing maximum singular value. The updated singular value matrix can be formulated as:
\begin{equation}\label{eq:low-rank_update}
        \mathbf{\Sigma}_{\zeta } = \mathrm{diag}(\max (\sigma _{j}-\zeta , 0)+\zeta \cdot \mathrm{step}(\sigma _{j}-\zeta )_{1\le j\le r}),
\end{equation}
where $\zeta =\mathrm{sigmoid}(\tau_{i})\cdot\bar{\sigma}$ is the threshold with $\bar{\sigma}$ being the maximum singular value in $\mathbf{\Sigma}$, and $\mathrm{step}(\cdot)$ denotes the Heaviside step function. The thresholded singular value matrix induces updated patches $\mathbf{x}_{i}=\mathbf{U}\mathbf{\Sigma}_{\tau}\mathbf{V}^{H}$. This operation guarantees the low-rank constraint by retaining only the singular values larger than the learned threshold. Once local thresholds for all patches are learned, the patches are unpatched to recover the original image size.

\paragraph{Image data consistency layer}
In the last update step in Eq.~\ref{eq:iter_sol_model}, we perform the data consistency step in the image domain using a gradient descent algorithm. Let us denote the original objective function in Eq.~\ref{eq:iter_sol} as $f(\mathbf{x})$. The derivation of $f(\mathbf{x})$ with respect to $\mathbf{x}$ is:
\begin{equation}
    \frac{\mathrm{d} f(\mathbf{x} )}{\mathrm{d} \mathbf{x}} = \mathbf{A}^{H}(\mathbf{Ax}-\mathbf{y}_{u})-\mu_{1}(\mathbf{p}_{n+1}-\mathbf{x})-\mu_{2}(\mathbf{q}_{n+1}-\mathbf{x}).
\end{equation}
We initialize the input to the DC layer as:
\begin{equation}\label{eq:dc_image_initialization}
    \mathbf{x}_{init} = \alpha \cdot \mathbf{p}_{n+1}+(1-\alpha) \cdot \mathbf{q}_{n+1}, \ \alpha=\frac{\mu_{1}}{\mu_{1}+\mu_{2}},
\end{equation}
so that the gradient is simplified to $\mathbf{A}^{H}(\mathbf{Ax}_{init}-\mathbf{y}_{u})$. By applying the gradient descent algorithm, the updated image of the DC layer is:
\begin{equation}\label{eq:dc_image_update}
    \mathbf{x}_{n+1} = \mathbf{x}_{init}-\eta \cdot (\mathbf{A}^{H}(\mathbf{Ax}_{init}-\mathbf{y}_{u})),  
\end{equation}
where $\eta$ is the trainable scaling parameter. Consequently, the data consistency layer takes the weighted combination of outputs from the image and low-rank subnetworks as input and induces fidelity to the sampled k-space $\mathbf{y}_{u}$ to update the image branch.

\subsubsection{K-space branch}\label{sec: ksp branch}
The k-space branch is dedicated to solving Eq.~\ref{eq:kspace_solution} and comprises a k-space subnetwork and a k-space data consistency layer. By introducing an auxiliary variable $\mathbf{r}$, Eq.~\ref{eq:kspace_solution} can be transformed into the unconstrained Lagrangian function using HQS:
\begin{equation}
    L_{\mu _{k}}(\mathbf{y},\mathbf{r}) = \frac{1}{2}\left \| \mathbf{My}-\mathbf{y}_{u}  \right \|_{2}^{2}+\lambda _{k}R_{k}(\mathbf{r} )+\frac{\mu _{k}}{2}\left \| \mathbf{r}-\mathbf{y}   \right \| _{2}^{2},       
\end{equation}
which can be solved iteratively:
\begin{equation}\label{eq:ksp_iter_sol}
    \left\{
    \begin{array}{l}
        \mathbf{r}_{n+1} = \arg\,\underset{\mathbf{r}}{\min}\ \lambda _{k}R_{k}(\mathbf{r} )+\frac{\mu _{k}}{2}\left \| \mathbf{r}-\mathbf{y}_{n}\right \| _{2}^{2}  \\
        \mathbf{y}_{n+1} = \arg\,\underset{\mathbf{y}}{\min}\ \frac{1}{2}\left \| \mathbf{My}-\mathbf{y}_{u}  \right \|_{2}^{2}+ \frac{\mu _{k}}{2}\left \| \mathbf{r}_{n+1}-\mathbf{y}\right \| _{2}^{2}.
    \end{array}\right.
\end{equation}

Similarly, we use a CNN, $\mathbf{K}_{n}$, operating in k-space to learn the generalized k-space prior (similar to a GRAPPA-based PI)~\citep{akccakaya2019scan} and perform data consistency in k-space to update $\mathbf{y}_{n+1}$. Then, Eq.~\ref{eq:ksp_iter_sol} can be formulated as learnable networks:
\begin{equation}
    \left\{
    \begin{array}{l}
        \mathbf{r}_{n} = \mathbf{K}_{n}(\mathbf{y}_{n-1}) \\
        \mathbf{y}_{n} = \mathbf{DC}_{\mathrm{K}, n}(\mathbf{r}_{n}),
    \end{array}\right.
\end{equation}

\begin{figure}[!t]
\includegraphics[width=\linewidth]{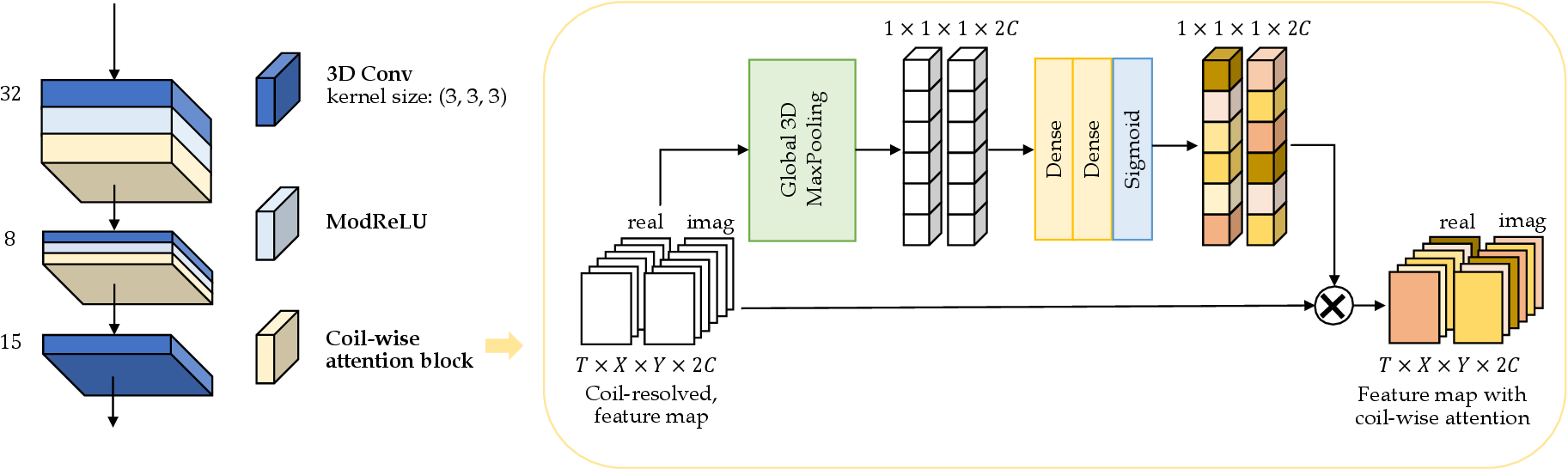}
\caption{K-space subnetwork with coil-wise attention blocks. A 3D three-layer convolutional neural network with attention blocks after each layer except the last one. The attention block squeezes and excites spatial-temporal features to generate an attention map along the MR coil dimension, assigning different weights to MR coils.}
\label{fig:ksp_network}
\end{figure}

\paragraph{K-space subnetwork}
The k-space subnetwork is designed to learn the generalized k-space prior. As depicted in Fig.~\ref{fig:ksp_network}, the k-space subnetwork is a complex-valued CNN with coil-wise attention blocks. Its structure closely resembles RAKI~\citep{akccakaya2019scan}, comprising a simple three-layer architecture. Unlike RAKI, however, we apply 3D convolutions in the spatial-coil dimensions and discard strides due to pseudo-random sampling patterns. The first two layers are followed by a ModReLU activation function and a coil-wise attention block. The last layer performs the output estimation with linear activation. All layers do not include bias terms, as the bias could adversely affect robustness when k-space undergoes linear scaling.

The attention mechanism is similar to the time-wise attention block in the image subnetwork. However, the dimensions on which it operates are different. In k-space, the real and imaginary parts are concatenated along the coil dimension, and the attention maps are calculated along this dimension. This modification stems from the fact that for multi-coil MR images, coil-resolved k-space data provides signal variations along the coils at distinct spatial locations, rendering the coil dimension more vital. Furthermore, the k-space network operates in the complex domain besides the attention block.

\paragraph{K-space data consistency layer}
The DC layer aims to align predicted values with actual observations and to solve the second problem in Eq.~\ref{eq:ksp_iter_sol}. This optimization function is convex and has an analytical solution:
\begin{equation}\label{eq:ksp_dc_solution}
    \left\{
    \begin{array}{lc}
        \mathbf{y}_{n+1}^{p} = \frac{\mathbf{y}_{u}^{p}+\mu _{k}\mathbf{r}_{n}^{p}}{1+\mu_{k}}  & p\in \Omega \\
        \mathbf{y}_{n+1}^{p} = \mathbf{r}_{n}^{p}  & p\notin \Omega, 
    \end{array}\right.
\end{equation}
where $\Omega$ represents the set of sampled k-space locations $p$. The solution means that, for the sampled points where the mask $\mathbf{M}^{p}=1$, the output of the DC is a linear combination of the acquired measurement and the predicted value controlled by the weighting parameter $\mu_{k}$. For non-sampled points where $\mathbf{M}^{p}=0$, the output of the DC layer corresponds to the predicted values from the previous k-space subnetwork.

\subsubsection{Information sharing layer}\label{sec: ISL}
In the problem formulation Sec.~\ref{sec: problem_formulation}, we have introduced the derivation of Eq.~\ref{eq:ISL_constrain}, which is implemented by the information sharing layer (ISL). The inputs to the ISL at the $n^{th}$ iteration are $\mathbf{x}_{n}$ and $\mathbf{y}_{n}$ from the image and k-space branches. To satisfy Eq.~\ref{eq:ISL_constrain}, they can be expressed as follows:
\begin{equation}\label{eq:ISL_update}
    \left\{
    \begin{array}{ll}
        \mathbf{y}_{n} = a\cdot (\mathbf{FSx}_{n})+(1-a)\cdot \mathbf{y}_{n}\\
        \mathbf{x}_{n} = b\cdot ((\mathbf{FS})^{H}\mathbf{y}_{n})+(1-b)\cdot\mathbf{x}_{n},
    \end{array}\right.
\end{equation}
where $a$ and $b$ are real numbers with $a,b \in \lbrack 0, 1\rbrack$. As previously explained, achieving perfect consistency between $\mathbf{x}_{n}$ and $\mathbf{y}_{n}$ can be challenging due to the inherent instability in the network training. Therefore, we set $a$ and $b$ in Eq.~\ref{eq:ISL_update} as trainable parameters, and Eq.~\ref{eq:ISL_update} becomes the update rule of the ISL.

In conclusion, a single iteration of the proposed A-LIKNet consists of (1) an image branch which includes an image subnetwork learning the sparse prior, a low-rank subnetwork enforcing local spatial-temporal low-rankness, and an image DC layer; (2) a k-space branch which is comprised of a k-space subnetwork learning the k-space regularization and a k-space DC layer; (3) an information sharing layer which exchanges and combines information from the two domains. Notably, the network weights are not shared across iterations, enabling each iteration to operate at different noise and artifact levels.

\section{Experiments}
\subsection{Database}
The dataset used in our experiments contains in-house 2D cardiac Cine, which was acquired on a 1.5T MRI scanner (MAGNETOM Aera, Siemens Healthineers, Erlangen, Germany) with a balanced steady-state free processing (bSSFP) sequence. The imaging protocol ensured left ventricular coverage, achieved through multiple consecutive breath-holds. The total scan time ranges between 122s and 611s, primarily falling within the range of 122s to 266s, depending on the number of slices and physiological characteristics of the patients. Data was recorded with flexible 18-channel body and 32-channel spine coils, resulting in 30 MR receiver channels. The sequence parameters are as follows: TE/TR=1.06/2.12ms, flip angle=52°, bandwidth=915 Hz/px, spatial resolution=1.9$\times$1.9mm$^2$, slice thickness=8mm. The acquired data has a temporal resolution of 40ms, encompassing 25 cardiac phases spanning a complete cardiac cycle. The database includes 129 subjects, amongst which are 38 healthy subjects and 91 patients with various cardiovascular diseases. We split the dataset into 115 subjects for training, including 34 healthy volunteers and 86 patients. The remainder of the dataset was used for testing. The study was approved by the local ethics committee, and all subjects gave written consent.

The undersampling masks used in all experiments are generated using a variable density incoherent spatiotemporal acquisition (VISTA) technique~\citep{ahmad2015variable}. Training data was generated by retrospective undersampling with VISTA. We normalized the magnitude of fully-sampled MR images to $\lbrack 0, 1\rbrack$. Coil sensitivity maps were estimated from the acquired auto-calibration signal data consisting of 24 central k-space lines using ESPiRIT and were compressed to 15 coils using the Berkeley Advanced Reconstruction (BART) toolbox~\citep{uecker2014espirit, uecker2016bart}. During training, the acceleration factors are randomly selected from $2\times$ to $24\times$. The undersampled k-space, zero-filled image, binary undersampling mask, and coil sensitivity maps form the inputs of A-LIKNet, and the fully-sampled k-space and image are used as labels in a supervised setting.

To compare the generalization performance, we further test the pre-trained networks on the prospectively undersampled data from the OCMR dataset~\citep{chen2020ocmr}. This prospective study does not involve training or fine-tuning. We directly test the pre-trained models of the retrospectively undersampled data on the OCMR dataset. The prospectively undersampled data used for testing were collected on a 1.5T MRI scanner (MAGNETOM Avanto, Siemens Healthineers, Erlangen, Germany) with a bSSFP sequence under free-breathing. More details about the dataset can be found in~\citep{chen2020ocmr}. 

\subsection{Implementation details}
The number of filters in the image sub-network starts from 12 for each UNet, which is doubled after the pooling layer in the encoder and halved after the up-sampling layer in the decoder. We use the 3D max pooling with pooling size 2 for the down-sampling operations and implement the up-sampling by a 3D transpose convolution with kernel size=1 and strides=2. For 2D spatial convolution, we employ a kernel size of 5$\times$5, while for temporal convolution and k-space convolution, the kernel size is set to 3. We employ the complex-valued ModReLU activation~\citep{arjovsky2016unitary} function as the nonlinear activation function.

The hyperparameters used in A-LIKNet are set as follows: the trainable low-rank threshold coefficient $\tau_{i}$ in Eq.~\ref{eq:low-rank_update} is initialized as $-2$ for each local spatial-temporal patch. For the patch-wise low-rank subnetwork, we choose the best performance patch size, which divides the total $T$=25 frames into $(n_{t}, n_{x}, n_{y})=(5,4,4)$ groups, implying that each spatial-temporal patch contains five frames with $1/16$ original spatial size. The trainable parameter $\alpha$ in Eq.~\ref{eq:dc_image_initialization} is initialized as $0.5$, meaning that the outputs of the image and the low-rank subnetworks have the same contribution to the DC layer at the start of training. Additionally, the trainable weights $\eta$ (Eq.~\ref{eq:dc_image_update}) and $\mu _{k}$ in the k-space data consistency layer (Eq.~\ref{eq:ksp_dc_solution}) are initialized as $1.0$. Furthermore, to set the same importance of the k-space and image branches at the start of training, we initialize the scaling factors $a$ and $b$ in the information sharing layer (Eq.~\ref{eq:ISL_update}) as $0.5$. The proposed A-LIKNet iterates $N$=8 times in total, resulting in 2,477,961 trainable parameters.

We train A-LIKNet in a supervised manner. During training, the network aims to minimize the pixel-wise mean absolute error (MAE) between the reconstructed k-space/image and the corresponding fully-sampled k-space/image. For complex-valued outputs, the loss function is calculated as follows:
\begin{equation}
    \begin{array}{ll}
        \mathcal{L}_{tot} &= \mathcal{L}_{\mathrm{I}}+\mathcal{L}_{\mathrm{K}} \\
        &= \frac{1}{P}\sum_{s = 1}^{S}  \sum_{p = 1}^{P}(\sqrt{d^{\ast}_{i,sp} \cdot d_{i,sp}}+\sqrt{d^{\ast}_{k,sp} \cdot d_{k,sp}}),
    \end{array} 
\end{equation}
where $\mathcal{L}_{\mathrm{I}}$ is the image branch loss, $\mathcal{L}_{\mathrm{K}}$ is the k-space branch loss, $d_{i,sp}=\mathbf{\hat{x}}_{sp}-\mathbf{x}_{\text{ref},sp} $ is the difference between the reconstructed image and the fully-sampled image at the $p^{th}$ pixel of the $s^{th}$ subject, and $d_{k,sp}=\mathbf{\hat{y}}_{sp}-\mathbf{y}_{\text{ref},sp}$ represents the k-space difference. The reconstructed image contains $P$ pixels, and there are $S$ subjects in the training dataset. 

We implemented the proposed A-LIKNet framework using \textit{Tensorflow}~\citep{abadi2016tensorflow} v2.5.0 with \textit{Keras}~\citep{chollet2021deep} v2.4.3. Complex-valued operations such as convolution and activation are implemented by MERLIN v0.3~\citep{merlin2022}. Networks were trained to converge (180 epochs) using an Adam optimizer~\citep{kingma2014adam} with a learning rate of $1\times$10$^{-4}$. Due to the high GPU memory burden, we set the batch size to one. The training of A-LIKNet took approximately 96 hours using 4 NVIDIA A6000 GPUs (48 GB VRAM). The source code is publicly available: \href{https://github.com/midas-tum/A-LIKNet}{https://github.com/midas-tum/A-LIKNet}.

All comparative and ablated experiments were conducted using the same VISTA sampling technique to ensure a fair comparison. For deep learning-based networks, the acceleration rate at each training step was randomly selected between $2\times$ and $24\times$. Both qualitative and quantitative evaluations were performed. For qualitative comparisons, all grey-scale images are normalized to $\lbrack 0, 1\rbrack$. For quantitative evaluations, we calculated the normalized root mean squared error (NRMSE), peak signal-to-noise ratio (PSNR), and structural similarity index measure (SSIM). The evaluation metrics were computed for all slices across all subjects in the test dataset. All networks were trained until convergence.

\subsection{Comparative methods}
We compare the proposed A-LIKNet to three conventional reconstruction methods: zero-filling, parallel-imaging compressed sensing reconstruction (PICS)~\citep{lustig2007sparse}, and k-t SLR~\citep{lingala2011accelerated}, as well as three deep learning-based methods: MoDL~\citep{aggarwal2018modl}, multi-domain KIKI-net~\citep{eo2018kiki}, and low-rank plus sparse network L+S-Net~\citep{huang2021deep}. 

For traditional methods, parameters were adjusted to achieve optimal performance. KIKI-Net, which was initially designed for 2D single-coil data, was adapted to our 2D+t multi-coil data to maintain a fair comparison and only focus on the multi-domain sharing aspect. We modified the original 2D convolutions to 3D convolutions, aligning them with the dimensionality of our A-LIKNet: spatial-coil in the k-space and spatial-temporal in the image domain. The data consistency layer applied the same gradient descent algorithm for the multi-coil data fidelity. Similarly, the original 2D convolutions in MoDL were adjusted to 3D convolutions to fit our Cine data. The iteration numbers of MoDL and L+S-Net matched those of A-LIKNet, iterating eight times. The remaining network structures, loss functions, training strategy, and other details remained consistent with the original implementations. The total trainable parameters and the final reconstruction time for a Cine image sequence are summarized in Tab.~\ref{tab:summary_comparison}.

\begin{table}[!t]
\centering
\caption{Summary of trainable parameters and reconstruction times for a complete 2D cardiac Cine dataset in all examined algorithms.}
\label{tab:summary_comparison}
\resizebox{0.6\textwidth}{!}{%
\begin{tabular}{cc|c|c}
\toprule
\multicolumn{2}{c|}{Methods}                                                                                           & Trainable params & Recon time (s) \\
\midrule
\multicolumn{1}{c|}{\multirow{2}{*}{\begin{tabular}[c]{@{}c@{}}conventional\\ methods\end{tabular}}}        & PICS     & None             & 26.66          \\
\multicolumn{1}{c|}{}                                                                                       & kt-SLR   & None             & 646.80         \\
\midrule
\multicolumn{1}{c|}{\multirow{4}{*}{\begin{tabular}[c]{@{}c@{}}deep learning-\\ based methods\end{tabular}}} & MoDL     & 339,980          & 13.74          \\
\multicolumn{1}{c|}{}                                                                                       & KIKI-Net & 479,352          & 12.94          \\
\multicolumn{1}{c|}{}                                                                                       & L+S-Net  & 262,671          & 4.94           \\
\multicolumn{1}{c|}{}                                                                                       & A-LIKNet & 2,477,961        & 42.16          \\
\bottomrule
\end{tabular}%
}
\end{table}

As shown in Tab.~\ref{tab:summary_comparison}, the proposed A-LIKNet has more trainable parameters than other networks due to multiple subnetworks. To ensure further fairness, we selected the best-performing and least-parameterized network, L+S-Net, from the comparative DL-based methods for an additional enlarged L+S-Net experiment. In this experiment, we augmented the original L+S-Net by increasing the spatial-temporal convolutional kernel size from (3,3,3) to (3,5,5), matching the kernel size in A-LIKNet. We summarized the hyperparameter changes in Tab.~\ref{tab:hyperparameter}. The enlarged L+S-Net  has 2,688,015 parameters, which is comparable to the proposed A-LIKNet.

\begin{table}[!t]
\centering
\caption{Summary of trainable parameters for original L+S-Net, enlarged L+S-Net, and the proposed A-LIKNet. The number of filters in A-LIKNet refers to the initial number in the UNet. (o: original, e: enlarged).}
\label{tab:hyperparameter}
\resizebox{0.58\textwidth}{!}{%
\begin{tabular}{c|c|c|c}
\toprule
\textbf{Hyperparameter}       & \textbf{L+S-Net (o)} & \textbf{L+S-Net (e)} & \textbf{A-LIKNet}    \\
\midrule
kernel size (s)        & 3                           & 3                          & 3                    \\ \midrule
kernel size (t)        & 3                           & 5                          & 5                    \\ \midrule
number of filters           & 32                          & 64                         & 12 (UNet) \\
\midrule
\textbf{Trainable params} & \textbf{262,671}            & \textbf{2,688,015}         & \textbf{2,477,961}   \\
\bottomrule
\end{tabular}%
}
\end{table}

\subsection{Ablations}
The different contributions of the image, k-space, and low-rank subnetworks were examined. To investigate the roles of the different modules within A-LIKNet, we conducted the following five ablation experiments: A-INet, A-KNet, A-LINet, A-IKNet, and LIKNet. The components included in each experiment are summarized in Tab.~\ref{tab:ablation_exps}. We furthermore tested the contribution of the attention mechanisms, marked as  'w/o' in Tab.~\ref{tab:ablation_exps} for the image or k-space subnetworks.

\begin{table}[ht]
\centering
\caption{Summary of network components for all ablation experiments. (y: yes, n: no, w\(/\)o: with\(/\)without the attention block).}
\label{tab:ablation_exps}
\resizebox{0.58\textwidth}{!}{%
\begin{tabular}{c|c|c|c|c|c|c}
\toprule
Exp      & $\mathbf{L}_{n}$ & $\mathbf{H}_{n}$ & $\mathbf{K}_{n}$ & $\mathbf{DC}_{I,n}$ & $\mathbf{DC}_{K,n}$ & ISL \\
\midrule
A-INet   & n                & y / w                        & n                           & y                   & n                   & n   \\ 
A-Knet   & n                & n                           & y / w                        & n                   & y                   & n   \\ 
A-LINet  & y                & y / w                        & n                           & y                   & n                   & n   \\ 
A-IKNet  & n                & y / w                        & y / w                        & y                   & y                   & y   \\ 
LIKNet   & y                & y / o                        & y / o                        & y                   & y                   & y   \\ 
A-LIKNet & y                & y / w                        & y / w                        & y                   & y                   & y   \\
\bottomrule
\end{tabular}%
}
\end{table}

Additionally, we conducted a series of ablation experiments with varying patch sizes in the low-rank subnetwork to explore the effect of the proposed patch-wise low-rank method. We use $(n_{t}, n_{x}, n_{y})$ to denote the number of patches in the temporal and spatial dimensions.

\section{Results}
\subsection{Reconstruction performance of A-LIKNet}
\begin{figure*}[ht]
\includegraphics[width=\linewidth]{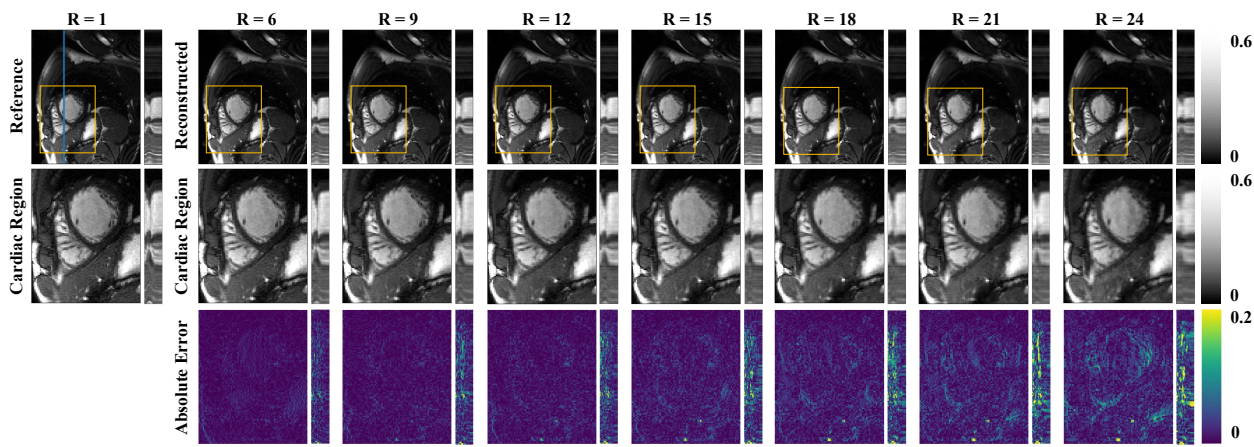}
\caption{Reconstructions in spatial (x-y) and spatial-temporal (y-t) plane of the proposed A-LIKNet for a retrospectively undersampled (VISTA) patient with transposition of the great vessels. Results for the acceleration factors R=6, 9, 12, 15, 18, 21, and 24 are shown in each column. The dynamic performance in the y-t plane corresponds to the blue line in the reference x-y plane image. The second row shows the enlarged views of the cardiac region (yellow box region). The third row presents the corresponding 5-times scaled absolute error maps.}
\label{fig:recon_results}
\end{figure*}

We first explored the performance of the proposed A-LIKNet at different acceleration rates. It is worth noting that since acceleration factors were randomly chosen between $2$ and $24$ during training, the pre-trained network can be used for testing at various accelerations without the need for retraining. The reconstruction results of A-LIKNet at different acceleration factors are depicted in Fig.~\ref{fig:recon_results}.

In Fig.~\ref{fig:recon_results}, we selected a frame from the diastolic phase of a representative patient in the test dataset. We show the reconstructions in x-y and y-t planes, along with magnified views of the cardiac region and the corresponding absolute error maps. Evidently, the proposed A-LIKNet effectively reconstructs the image content and recovers dynamic information even at different acceleration rates. Up to $12\times$ acceleration, only minor errors are visible in the $5\times$ magnified error maps, demonstrating excellent reconstruction performance. With the increase in acceleration, although there exists a slight increase in errors at $24\times$ acceleration, the reconstructed morphology and details remained acceptable. We observed a similar performance throughout the complete test dataset. Furthermore, we attempted to test the pre-trained network on $30\times$ accelerated data to test it out of the domain. Even though the network had only seen images with a maximum acceleration of $24\times$ during training, the A-LIKNet could still remove most artifacts and reconstruct the image structure, as shown in Fig.~\ref{fig:recon_30x} in the supplementary material.

\begin{figure*}[!t]
\includegraphics[width=\linewidth]{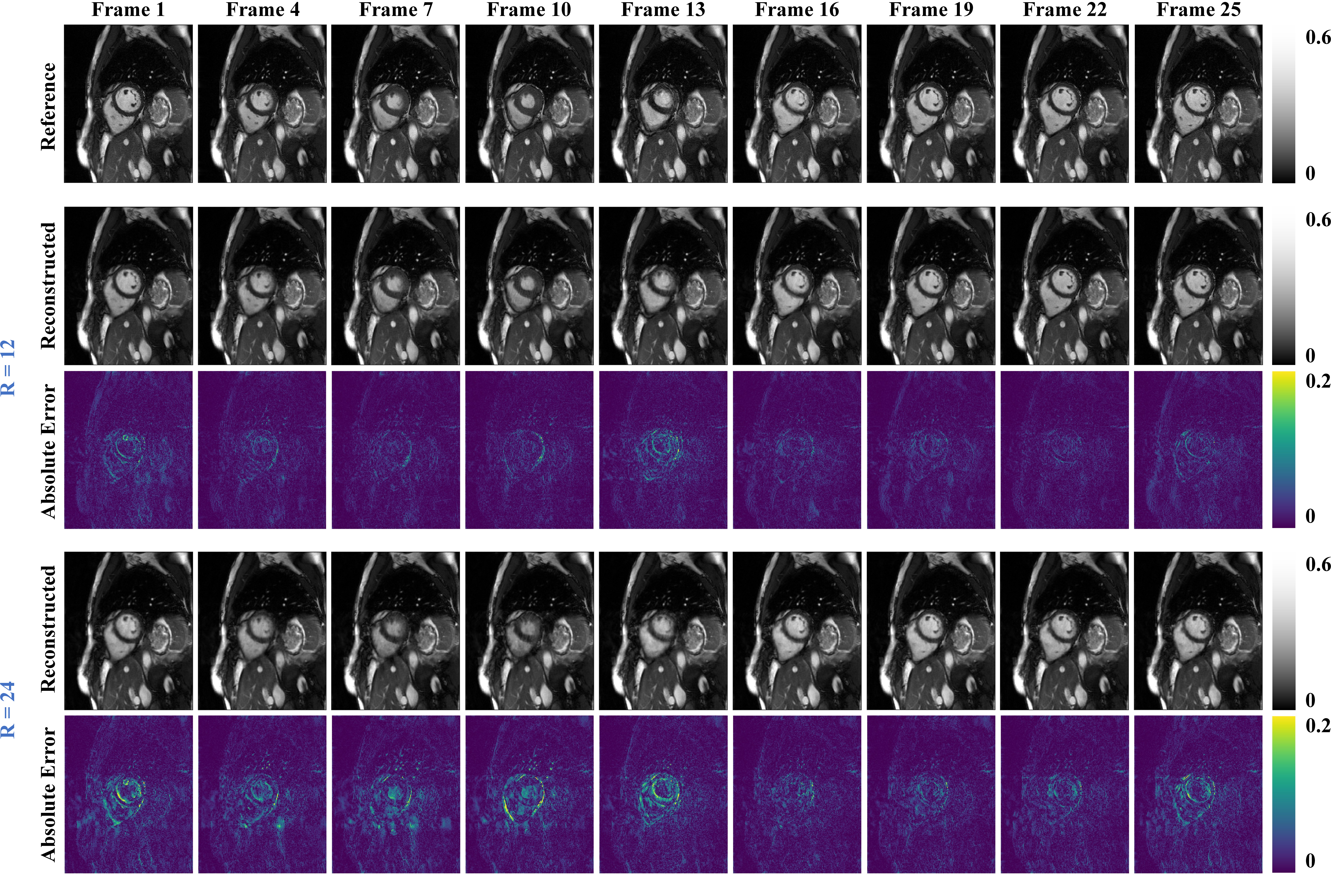}
\caption{Reconstructions in spatial (x-y) plane of the proposed A-LIKNet for a retrospectively undersampled (VISTA) healthy subject. Results for every 3rd frame over one cardiac cycle are shown in each column. Both R=12 and R=24 reconstructions are shown alongside the corresponding absolute error maps. The enlarged views of the cardiac region are shown in Fig.~\ref{fig:Recon_result_cycle_cardiac} in the supplementary material.}
\label{fig:recon_results_cycle}
\end{figure*}

To provide a more intuitive representation of the reconstruction performance on dynamic images, we showcase the reconstructions of A-LIKNet for an entire cardiac cycle of a healthy volunteer from the test dataset in Fig.~\ref{fig:recon_results_cycle}. We selected every third frame for display. Overall, the reconstruction performance of the diastolic phase (frames 16 to 25) is superior to that of the systolic phase (frames 1 to 13). This performance decrease is likely due to the rapid motion of the heart during the systolic phase, leading to larger inter-frame displacements and increased reconstruction challenges.

In general, A-LIKNet presents convincing reconstruction performance for different accelerations and time frames. The quantitative analysis in Tab.~\ref{tab:ablation_quant} shows that A-LIKNet achieves a high structural similarity score of nearly 0.9 at $24\times$ acceleration, which is an average value calculated on all test subjects across all slices and time frames, demonstrating that our network is fully capable of performing reconstructions at such a high acceleration rate.

\subsection{Comparative studies}
\subsubsection{Qualitative evaluation}
\begin{figure*}[!t]
\includegraphics[width=\linewidth]{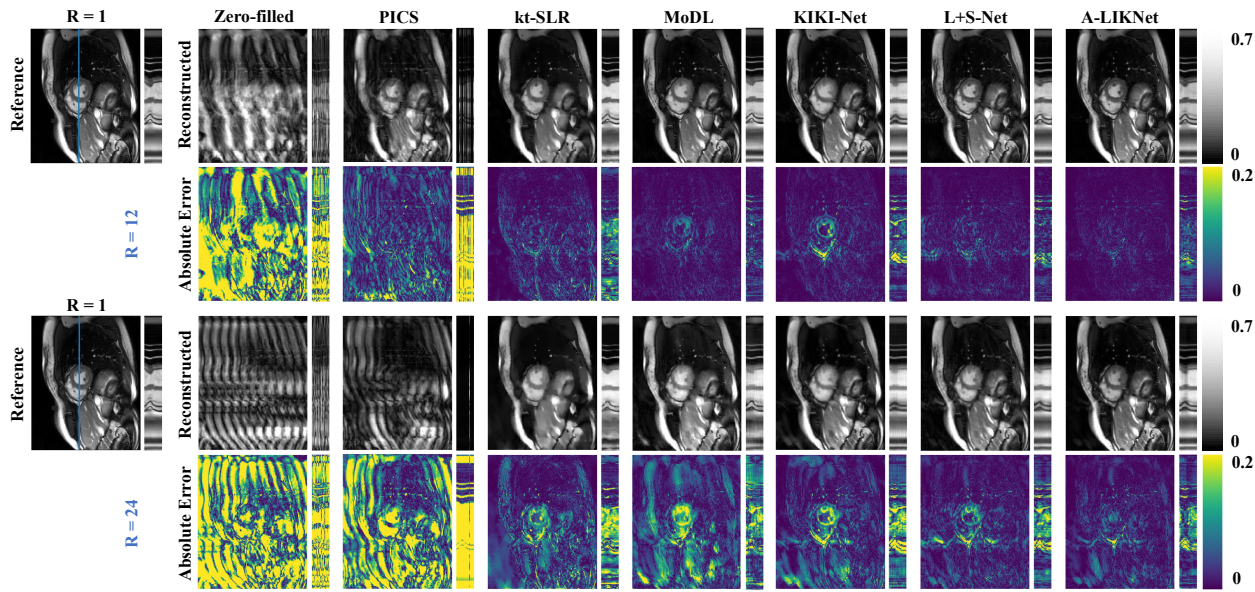}
\caption{Reconstructions in spatial (x-y) and spatial-temporal (y-t) plane of the proposed A-LIKNet in comparison to zero-filled, PICS, kt-SLR, MoDL, KIKI-Net, and L+S-Net for a patient with active myocarditis who was retrospectively undersampled with VISTA sampling. The dynamic performance in the y-t plane corresponds to the blue line in the reference x-y plane image. Both R=12 (top) and R=24 (bottom) reconstructions are shown alongside the corresponding absolute error maps. The enlarged views of the cardiac region are shown in Fig.~\ref{fig:compare_results_cardiac} in the supplementary material.}
\label{fig:compare_results}
\end{figure*}

Fig.~\ref{fig:compare_results} displays the reconstruction results for $12\times$ and $24\times$ retrospectively undersampled Cine images of another patient in the test dataset, distinct from the one shown in Fig.~\ref{fig:recon_results}. We compare our A-LIKNet to zero-filling, PICS~\citep{lustig2007sparse}, k-t SLR~\citep{lingala2011accelerated}, and deep learning-based methods including MoDL~\citep{aggarwal2018modl}, KIKI-net~\citep{eo2018kiki}, L+S-Net~\citep{huang2021deep}. A frame from the systolic phase is shown, as cardiac motion is the largest in this phase, making the reconstruction more challenging to capture the cardiac dynamics. Similarly, Fig.~\ref{fig:compare_results} illustrates the reconstructions in the x-y and y-t planes, along with the absolute error maps.

From Fig.~\ref{fig:compare_results}, it is apparent that the traditional PICS method struggles with the VISTA sampling pattern. PICS fails to reconstruct the $24\times$ undersampled images and exhibits significant artifacts even at $12\times$ acceleration. Another traditional approach, kt-SLR, achieves acceptable results at $12\times$ acceleration but shows noticeable blur at edges and over-smoothing of artifacts at $24\times$ acceleration. Among the deep learning-based methods, both MoDL and KIKI-Net exhibit significant errors in the myocardial region. While L+S-Net performs better than these two methods, some residual ripple-like artifacts are still observable in the reconstructed images at $24\times$ acceleration. In contrast, whether under a moderate $12\times$ or an extremely high $24\times$ acceleration, the proposed A-LIKNet outperforms all other methods, showcasing significantly fewer errors, a more precise depiction of the myocardium, and superior contrast preservation. Furthermore, it is worth noting that A-LIKNet can achieve effective reconstructions for other body parts, such as the liver and lungs, in addition to the cardiac region. This performance demonstrates a generalization capability of A-LIKNet across different tissues. It effectively learns to remove artifacts rather than memorizing the structure of a specific organ.

From the spatial-temporal (y-t) plane images in Fig.~\ref{fig:compare_results}, it is evident that traditional methods exhibit more errors in reconstructing dynamic features than DL-based methods. It is challenging for PICS to capture spatial-temporal information, while kt-SLR can recover the general structure but still exhibits issues like over-smoothing and noticeable error maps. Their unsatisfactory performance stems from being typically based on static models, which hinders their ability to capture temporal information when dealing with dynamic images. In contrast, the proposed A-LIKNet demonstrates the least error at $24\times$ acceleration compared to all other methods, indicating that the network can effectively learn dynamic information. Several crucial structures within the network contribute to the superior performance. The spatial-temporal convolutions in the image sub-network contribute to capturing spatial-temporal correlations, while the separate 2D+t convolutions further improve the independence of learning in the temporal domain. Including the low-rank sub-network enhances dynamic behavior modeling, aiding the network in better learning and representing key features of the dynamic information. Furthermore, the attention mechanism in the image sub-network allows the network to assign greater weights to more informative frames adaptively.

The performance of the enlarged L+S-Net of the same subject is compared to the original L+S-Net and the proposed A-LIKNet in Fig.~\ref{fig:LSNet_enlarged} in the supplementary material. We observed that the increase in parameters in the L+S-Net did not result in a noticeable improvement, and the A-LIKNet still outperforms the L+S-Net with a similar amount of parameters.

\subsubsection{Quantitative evaluation}
\begin{figure*}[!t]
\includegraphics[width=\linewidth]{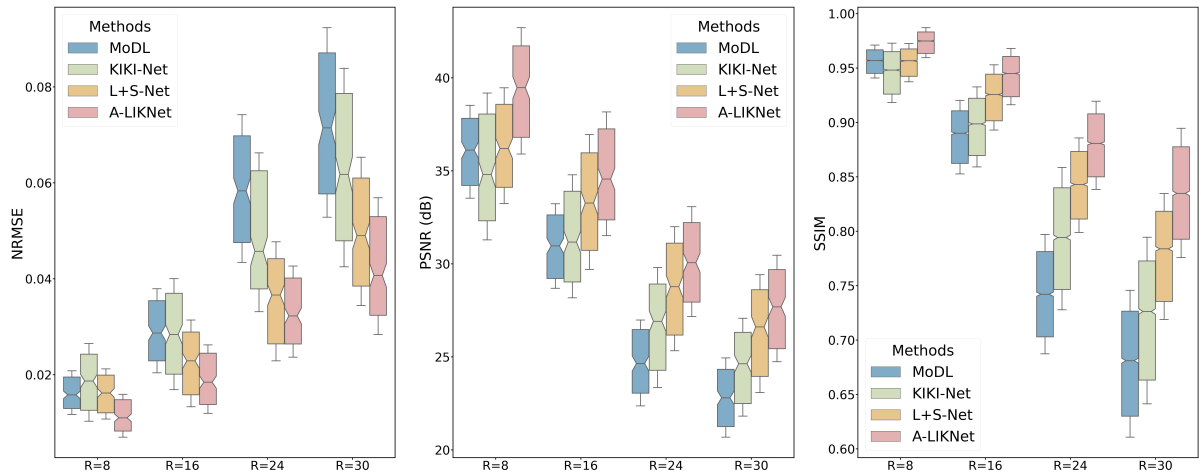}
\caption{Quantitative analysis in terms of NRMSE, PSNR, and SSIM between MoDL, KIKI-Net, L+S-Net, and the proposed A-LIKNet reconstruction. Each metric was computed for accelerations R=8, 16, 24, and 30. Results are calculated for all subjects in the test dataset and depicted in box plots (horizontal line: median, box: 25\% and 75\% percentile, whiskers: 0.2 $\cdot$ interquartile range).}
\label{fig:compare_results_quant}
\end{figure*}

To provide a fair comparison of the performance of various deep learning-based methods, we reconstructed Cine images for all slices across all subjects in the test dataset. We calculated the NRMSE, PSNR, and SSIM between the reconstruction results and the fully-sampled images for four different acceleration factors: $8\times$, $16\times$, $24\times$, and $30\times$. Note that during training, the maximum acceleration factor is $24\times$, and we validate different DL-based methods on $30\times$ to test their generalization ability. VISTA sampling masks were randomly generated for each slice during the validation process. The final results are presented as box plots in Fig.~\ref{fig:compare_results_quant}.

The results of the quantitative evaluation are in line with the qualitative comparisons. As seen in Fig.~\ref{fig:compare_results_quant}, the proposed A-LIKNet outperforms the other deep learning-based reconstruction methods for all quantitative evaluation metrics under all acceleration rates. In terms of SSIM, which can better emulate human perception, A-LIKNet demonstrates notable performance with average SSIM scores of 0.97, 0.94, and 0.88 at $8\times$, $16\times$, and $24\times$ accelerations, respectively. In contrast, the average SSIM values for L+S-Net decrease to 0.95, 0.92, and 0.83. KIKI-Net displays inferior performance compared to A-LIKNet and L+S-Net, with average SSIM values of 0.94, 0.89, and 0.78, while MoDL exhibits similar performance with average SSIM values of 0.95, 0.88, and 0.74. Additionally, we conducted a Wilcoxon rank-sum test to compare the differences between the quantitative evaluation data. The results indicate that for each evaluation metric at every acceleration level, the p-values between A-LIKNet and other methods are significantly below 0.05, demonstrating a significant difference in quantitative results. Furthermore, the mean SSIM value of the images reconstructed by A-LIKNet at $30\times$ acceleration dropped by $14.6\%$ compared to that at $8\times$ acceleration. In contrast, the performance of MoDL, KIKI-Net, and L+S-Net declined by $29.0\%$, $23.9\%$, and $18.7\%$, respectively, when pushing the acceleration from $8\times$ to $30\times$. This difference indicates that the A-LIKNet demonstrates better stability and generalization as acceleration increases. Also, the small percentiles (box area) and standard deviations (whiskers) indicate that A-LIKNet has low variance in reconstructions across all subjects in the test dataset, indicating good robustness.

\subsubsection{Prospective study}
\begin{figure*}[ht]
\includegraphics[width=\linewidth]{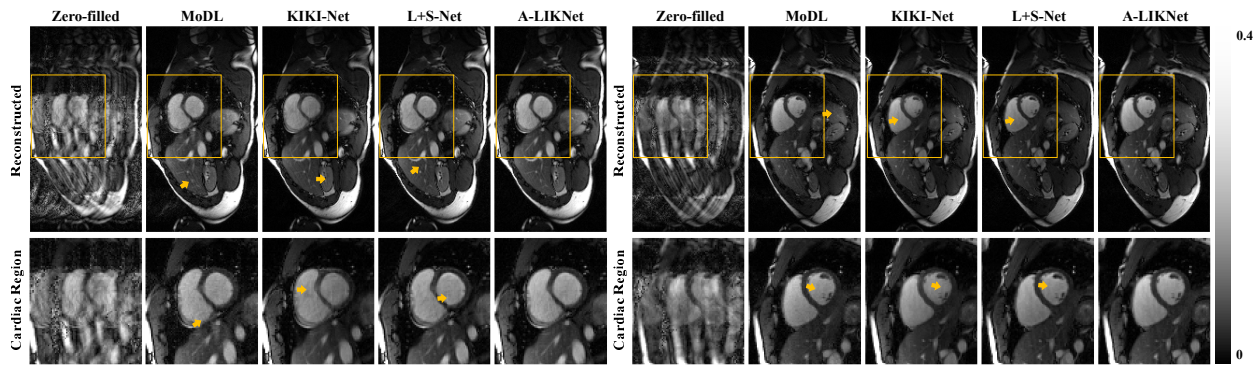}
\caption{Reconstructions in spatial (x-y) plane of the proposed A-LIKNet in comparison to zero-filled, MoDL, KIKI-Net, and L+S-Net for a healthy subject who was prospectively undersampled with VISTA sampling (R=8). Two different slices are shown on the right and left sides. The second row shows the enlarged views of the cardiac region (yellow box region).}
\label{fig:prosp_result}
\end{figure*}

To investigate the generalization performance of different DL-based networks, we tested pre-trained networks on the prospectively real-time undersampled OCMR dataset~\citep{chen2020ocmr}. It is important to note that the pre-trained networks were not fine-tuned on the OCMR dataset but were used directly for testing. Fig.~\ref{fig:prosp_result} displays two $8\times$ undersampled short-axis slices and the reconstruction results of MoDL, KIKI-Net, L+S-Net, and the proposed A-LIKNet.

In general, all four deep learning networks can reconstruct the prospectively undersampled images with high quality. Although the differences are subtle, A-LIKNet performs slightly better than the other methods. In both the full-view and magnified cardiac region images, we can observe that MoDL, KIKI-Net, and L+S-Net still exhibit some residual stripe artifacts (indicated by yellow arrows). Additionally, the reconstructed images of A-LIKNet display better contrast.

\subsection{Ablation studies}
\begin{figure*}[ht]
\includegraphics[width=\linewidth]{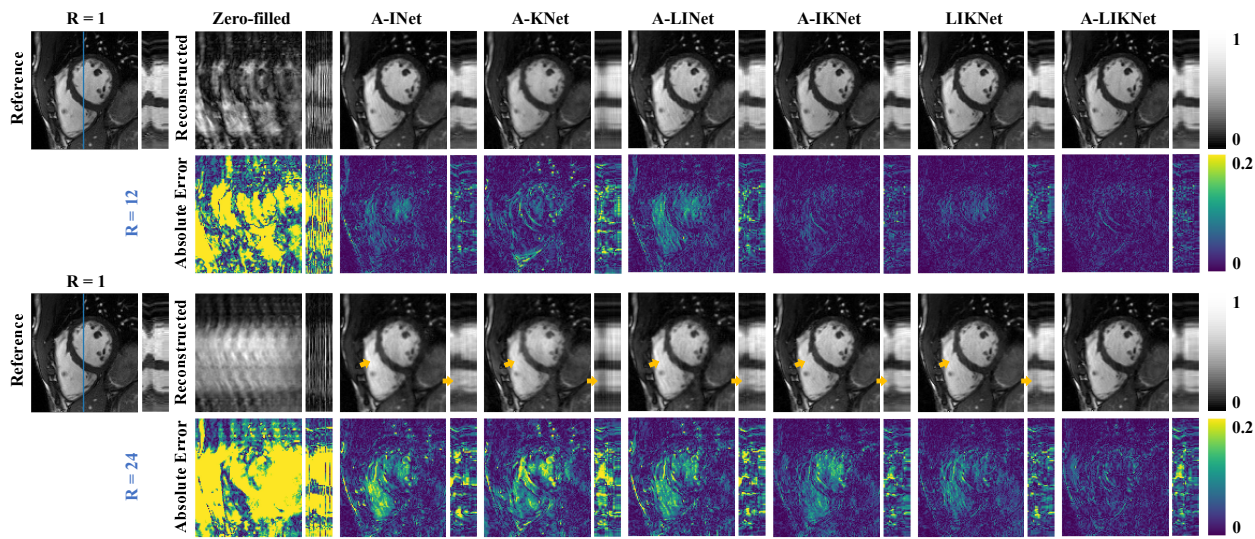}
\caption{Ablations: Reconstructions in spatial (x-y) and spatial-temporal (y-t) plane of the proposed A-LIKNet and ablation experiments for a healthy subject who was retrospectively undersampled with VISTA sampling. The abbreviations for each ablation experiment and the network configurations can be found in Tab.~\ref{tab:ablation_exps}. The dynamic performance in the y-t plane corresponds to the blue line in the reference x-y plane image. Both R=12 (top) and R=24 (bottom) results are shown alongside the corresponding absolute error maps.}
\label{fig:ablation_results}
\end{figure*}

To assess the role of each component in the proposed A-LIKNet, we conducted a series of ablation experiments as listed in Tab.~\ref{tab:ablation_exps}. Fig.~\ref{fig:ablation_results} displays the reconstruction results of various networks for a representative healthy subject from the test dataset. A frame from the diastolic phase is depicted, but similar observations also apply to other cohorts, accelerations, and frames. Quantitative metrics (NRMSE, PSNR, and SSIM) were also evaluated for all subjects in the test dataset at three different acceleration factors ($8\times$, $16\times$ and $24\times$), and the results are summarized in Tab.~\ref{tab:ablation_quant}.

\begin{table}[!t]
\centering
\caption{Quantitative evaluation of ablation experiments: The average and standard deviation of NRMSE, PSNR and SSIM in the reconstructed images for all subjects in the test dataset at R=8, R=16 and R=24 (mean±std). The asterisk(*) denotes statistically significant differences (p $<$ 0.05) between each method and A-LIKNet, as calculated by the Wilcoxon rank sum test.}
\label{tab:ablation_quant}
\resizebox{0.7\textwidth}{!}{%
\begin{tabular}{c|cccc}
\toprule
Accelerations               & Methods  & NRMSE ($\times10^{-2}$)     & PSNR (dB)            & SSIM               \\
\midrule
\multirow{5}{*}{R=8}  & A-INet   & 1.55±0.52$^{*}$             & 36.74±3.01$^{*}$    & 0.95±0.02$^{*}$    \\
                            & A-KNet   & 3.40±1.28$^{*}$             & 29.93±3.23$^{*}$    & 0.90±0.04$^{*}$    \\
                            & A-LINet  & 1.63±0.55$^{*}$             & 36.32±3.03$^{*}$    & 0.95±0.02$^{*}$    \\
                            & A-IKNet  & 1.29±0.49 \qquad            & 38.48±3.33 \qquad   & 0.95±0.02$^{*}$    \\
                            & LIKNet   & 1.26±0.47 \qquad            & 38.65±3.23 \qquad   & \textbf{0.97±0.02}$^{*}$ \\
                            & A-LIKNet & \textbf{1.22±0.48} \qquad   & \textbf{38.99±3.37} \qquad & \textbf{0.97±0.02} \qquad \\
\midrule
\multirow{5}{*}{R=16} & A-INet   & 2.63±0.82$^{*}$             & 32.17±2.82$^{*}$    & 0.90±0.04$^{*}$    \\
                            & A-KNet   & 4.80±1.66$^{*}$             & 26.85±3.11$^{*}$    & 0.81±0.05$^{*}$    \\
                            & A-LINet  & 2.66±0.85$^{*}$             & 31.85±2.82$^{*}$    & 0.89±0.04$^{*}$    \\
                            & A-IKNet  & 2.15±0.77 \qquad            & 34.07±3.19 \qquad   & 0.93±0.03$^{*}$    \\
                            & LIKNet   & 2.05±0.74 \qquad            & 34.27±3.18 \qquad   & 0.93±0.03$^{*}$    \\
                            & A-LIKNet & \textbf{1.99±0.73} \qquad   & \textbf{34.60±3.21} \qquad & \textbf{0.94±0.03} \qquad \\
\midrule
\multirow{5}{*}{R=24} & A-INet   & 4.75±1.47$^{*}$             & 27.30±2.75$^{*}$     & 0.79±0.06$^{*}$    \\
                            & A-KNet   & 6.91±2.38$^{*}$             & 23.64±3.29$^{*}$     & 0.71±0.07$^{*}$    \\
                            & A-LINet  & 4.55±1.45$^{*}$             & 27.33±2.64$^{*}$     & 0.79±0.06$^{*}$    \\
                            & A-IKNet  & 3.61±1.25 \qquad            & 29.36±3.34$^{*}$     & 0.86±0.05$^{*}$    \\
                            & LIKNet   & 3.67±1.31 \qquad            & 29.51±2.93 \qquad    & 0.86±0.05$^{*}$    \\
                            & A-LIKNet & \textbf{3.39±1.14} \qquad   & \textbf{30.11±3.18} \qquad & \textbf{0.88±0.04} \qquad \\
\bottomrule
\end{tabular}%
}
\end{table} 

\subsubsection{The effect of parallel branches}
Compared to existing methods, one crucial innovation of the proposed A-LIKNet is its parallel-branch structure, where two branches are dedicated to learning information separately in the k-space and image domain. The ablation experiments A-INet, A-KNet, and A-LINet are single-domain networks, either learning in the image domain or the k-space with a single branch. In contrast, multi-domain networks like A-IKNet, LIKNet, and A-LIKNet have parallel branches and information sharing layers between them.

By comparing the reconstruction results in Fig.~\ref{fig:ablation_results}, we can observe that the reconstructed image quality of single-domain networks is worse than that of multi-domain networks, the distinction particularly noticeable in the error maps. Multi-domain networks with the parallel-branch structure not only independently learn information in the frequency and image domains but also exchange information in the ISL, preventing the single branch from getting stuck in a local minima or saddle point. The network benefits from the proposed parallel-branch architecture and information sharing layers, maximizing the utilization of sampling information and thus better removing aliasing artifacts from the images.

The quantitative analysis in Tab.~\ref{tab:ablation_quant} also reflects the advantages of the parallel-branch structure. At the milder acceleration factor of $8\times$, multi-domain networks slightly outperform single-domain networks, but as the acceleration factor increases, the advantage of the multi-domain becomes more evident. At $24\times$ acceleration, multi-domain networks exhibit significantly smaller NRMSE, higher PSNR, and higher SSIM, highlighting their superiority.

\subsubsection{The effect of k-space branch}
The impact of the k-space branch can be explored through a comparison between A-LIKNet and A-LINet. From Fig.~\ref{fig:ablation_results}, we observed that the reconstruction results of A-LINet contain residual oscillatory artifacts that are even being sharpened. At $24\times$ acceleration, including the k-space branch helps remove aliasing artifacts, resulting in a more precise depiction of details, such as the edges of papillary muscles, compared to the relatively blurry boundaries of A-LINet. Regarding dynamic information learning, the y-t images of A-LINet exhibit noticeable noise and fail to reconstruct the temporal dynamics. Moreover, introducing k-space learning has a positive impact on recovering contrast information. The A-LIKNet reconstruction results present a contrast closest to the reference images. Quantitative metrics also support a consistent conclusion where A-LIKNet outperforms A-LINet significantly in all evaluation metrics at different acceleration factors.

Furthermore, the impact of the k-space branch can be deduced by comparing A-INet and A-IKNet. In the x-y plane, whether in terms of artifact removal or edge sharpness, the network with the k-space branch outperforms the ones working solely in the image domain. Dynamic information learning in the y-t plane is also improved by the participation of k-space information. Quantitative metrics corroborate this finding.

From both comparisons, joining the k-space branch with the image branch facilitates context learning. While k-space contains redundant information compared to the image domain, convolutions in the frequency domain can compensate for the limited local field of view of image domain convolutions. Furthermore, k-space harbors valuable coil-resolved information, which differs from the image domain where we typically work with coil-combined images.

\subsubsection{The effect of low-rank subnetwork}
By comparing the proposed A-LIKNet with A-IKNet, we can observe the role of the low-rank subnetwork. At $12\times$ acceleration, both networks exhibit comparable reconstruction results. However, as the acceleration factor increases, A-IKNet shows significantly more error at $24\times$ acceleration. Since there are no evident visual differences in the reconstructed images, the errors appear to be inaccuracies in signal intensities. Due to the aggressive acceleration, artifacts introduced in the undersampled images make it challenging to learn contrast information accurately. However, thanks to the inclusion of low-rank constraints that suppress the noise, the proposed A-LIKNet is able to learn signal intensity correctly while removing artifacts.

The quantitative analysis indicates that the inclusion of the low-rank subnetwork leads to a slight improvement in reconstruction performance, with this improvement becoming more pronounced as the acceleration factor increases. Furthermore, in most cases, A-LIKNet exhibits smaller variances in metrics than A-IKNet, demonstrating the added robustness of the low-rank subnetwork against noise influence.

In addition, the effect of the patch-wise operation in the low-rank subnetwork of A-LIKNet is qualitatively compared in Fig.~\ref{fig:ablation_patch} and quantitatively evaluated in Fig.~\ref{fig:patch_quant} in the supplementary material. In Fig.~\ref{fig:ablation_patch}, we compared the proposed A-LIKNet with (5,4,4) patch division to the ablated experiment without patch division. We observed that the network with patch division demonstrates superior reconstruction quality at both $12\times$ and $24\times$ accelerations, affirming the efficacy of learning low-rank characteristics for each spatial-temporal patch, as adjacent temporal frames and spatial pixels exhibit stronger correlations. Quantitative metrics are calculated for varying spatial patch sizes and presented as box plots in Fig.~\ref{fig:patch_quant}. We observed that moderate patch numbers yield the best performance, striking a balance between containing sufficient relevant information without overwhelming the network with irrelevant details. However, variations in patch division numbers have minimal impact on the network performance. Therefore, we selected the (5,4,4) patch division, which showed the best performance in terms of both mean and variance, for comparison to other methods.

\subsubsection{The effect of attention mechanism}
Another innovation in the proposed A-LIKNet involves incorporating attention mechanisms in both k-space and image subnetworks. The impact of attention can be investigated by comparing A-LIKNet with LIKNet. From Fig.~\ref{fig:ablation_results}, we observed that LIKNet has the best reconstruction performance in all ablation experiments. However, compared to A-LIKNet, which incorporates attention mechanisms, there are still minor discrepancies. In the $24\times$ reconstructed image of LIKNet, oscillatory artifacts are still present in the right ventricle (indicated by yellow arrows). The edge and the contrast of the papillary muscles are less precise than in A-LIKNet. 

The quantitative evaluation is consistent with our observations. At lower acceleration factors, the metrics for A-LIKNet and LIKNet are comparable. However, the attention mechanism further enhances the reconstruction quality at higher acceleration factors. In conclusion, allowing the network to autonomously select more important coil information in k-space and learn more critical time frames in the image domain has a positive impact on improving network performance.

\subsection{Trainable parameters}
\begin{figure}[t]
\centering
\includegraphics[width=0.7\textwidth]{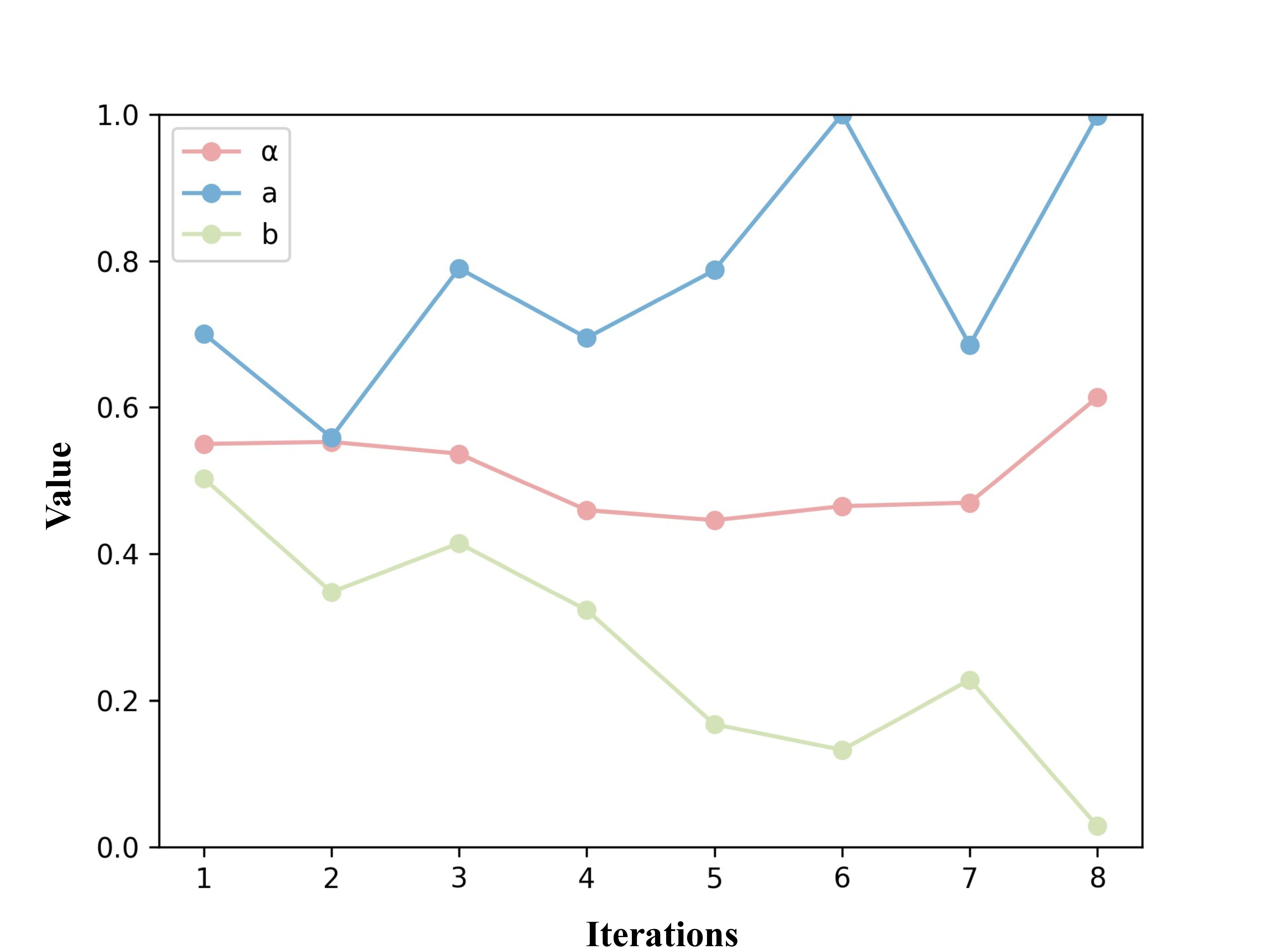}
\caption{Trained values of trainable parameters $\alpha$ in Eq.~\ref{eq:dc_image_initialization}, $a$ and $b$ in Eq.~\ref{eq:ISL_update}. The horizontal axis represents the iteration module within the unrolled network, and the vertical axis denotes the final values after training.}
\label{fig:trainable_paras}
\end{figure}

In the proposed A-LIKNet, some hyperparameters have been set as trainable to enhance model performance and ensure generalization ability. Observing the final values of these parameters allows us to draw interesting conclusions and insights. Fig.~\ref{fig:trainable_paras} summarizes the final values of the trainable parameters $\alpha$ in Eq.~\ref{eq:dc_image_initialization}, $a$ and $b$ in Eq.~\ref{eq:ISL_update}, with the curves depicting the trends of these parameters over iterations in the unrolled network. The initial values of these parameters were all set to 0.5 in each iteration.

In the image domain, the parameter $\alpha$ in Eq.~\ref{eq:dc_image_initialization} determines the weights between the low-rank and sparse priors. By observing $\alpha$ in all eight iterations in Fig.~\ref{fig:trainable_paras}, we found that the initially set value of 0.5 eventually settles in the range of 0.4 to 0.6. This suggests that the low-rank and sparse regularization terms contribute similarly during training, which also confirms the necessity of considering both priors for dynamic image reconstruction. Our local spatial-temporal low-rank subnetwork can effectively leverage the low-rank nature of dynamic images, complementing the image subnetwork in image domain learning.

In the information sharing layer, the weights of information from the k-space and image branches are determined by the parameters $a$ and $b$ in Eq.~\ref{eq:ISL_update}. A smaller $a$ indicates that the reconstruction result from the k-space branch influences more the update of k-space values in the ISL. Similarly, a smaller $b$ means that the image branch mainly determines the image update in ISL. As shown in Fig.~\ref{fig:trainable_paras}, the values of $a$ in all iterations are found to be greater than 0.5, while $b$ are all less than 0.5. This indicates that, for both the image and k-space updates, the image branch has a more significant contribution to information sharing. As the iterations progress and get closer to the output, the weight assigned to the image domain becomes more prominent. In the final iteration, $a$ is greater than 0.9, and $b$ is smaller than 0.1. To understand this phenomenon, we can look at the results of the ablation experiments comparing A-LINet and A-KNet. Both qualitative and quantitative analyses show that A-LINet, the image branch in A-LIKNet, outperforms A-KNet, the k-space branch in A-LIKNet. The better performance explains why the image domain information is dominant: learning in the image domain provides more assistance for image reconstruction.

\section{Discussion}
In this work, we developed a novel attention-incorporated network for sharing low-rank, image, and k-space information during MR image reconstruction (A-LIKNet) to achieve single breath-hold cardiac Cine imaging. A-LIKNet consists of three components: the k-space branch, the image branch, and the information sharing layers (ISL). We proposed a novel parallel-branch architecture that enables independent learning in the k-space and image domain. With the ISL connecting the two parallel branches, the utilization of multi-domain information is maximized. In the image branch, we do not only exploit the sparse priors but also leverage the intrinsic low-rank priors of dynamic images. We designed a local spatial-temporal low-rank module to adaptively enforce low-rank constraints for each spatial-temporal patch. Additionally, we introduced attention mechanisms in the network, allowing the k-space sub-network to assign different weights to the features from different coils, while the image sub-network can focus on the more crucial temporal frames.

\subsection{Experimental summaries}
The reconstruction results of A-LIKNet shown in Fig.~\ref{fig:recon_results} and Fig.~\ref{fig:recon_results_cycle} demonstrate that the trained network can effectively reconstruct images at different accelerations and time frames. We observed that while the trained A-LIKNet can handle image reconstruction under different conditions, the reconstruction performance slightly deteriorates during the cardiac systolic phase because the rapid motion of the myocardium leads to larger displacements between adjacent time frames, reducing the available redundant spatial-temporal information for sharing. Additionally, the greater number of frames in the diastolic phase compared to the systolic phase creates an unavoidable dataset imbalance, which may contribute to better performance during the diastolic phase. Furthermore, we observed that the reconstruction results for the first and last frames are slightly inferior to the rest. We speculate that this might be due to the absence of half of the spatiotemporal information during the learning process for the first and last frames.

We compared the A-LIKNet to various MR reconstruction methods, including traditional approaches and deep learning-based networks. Both qualitative (Fig.~\ref{fig:compare_results}) and quantitative results (Fig.~\ref{fig:compare_results_quant}) indicate that the proposed A-LIKNet outperforms other methods regarding artifact removal, dynamic characteristics learning, and details depiction. It also exhibits slightly better generalization performance in the prospective study (Fig.~\ref{fig:prosp_result}).

The excellent performance of A-LIKNet and the roles of various modules within the network can be elucidated through a series of ablation experiments we conducted. We found that single-domain networks with a single branch (A-INet, A-KNet, and A-LINet) exhibited obviously more errors compared to multi-domain networks with parallel branches (A-IKNet, LIKNet, A-LIKNet) (Fig.~\ref{fig:ablation_results}). The reasons can be summarized in the following two points: Firstly, the parallel network structure ensures the independence of learning in the k-space and image domains. Secondly, the ISL maximizes the utilization and exchange of multi-domain information, which may prevent the network from getting stuck in local optima within a single domain. While the k-space and image domain contain redundant information, the join of the k-space branch enhances the network performance (as seen in the comparison between A-LINet and A-LIKNet). This improvement is due to the limited receptive field of view of convolutional kernels in the image subnetwork, whereas k-space convolutions effectively utilize information from all pixels in the image domain. Furthermore, unlike the coil-combined image used in the image domain, k-space data contains valuable coil-resolved information.

Observing the single-domain networks, we noticed that the image network (A-INet) outperforms the k-space network (A-KNet), exhibiting more apparent structures and sharper details. The superior performance of the image network is also reflected in the trainable parameters of the ISL (Fig.~\ref{fig:trainable_paras}), where information coming from the image branch tends to dominate the update of both the k-space and image branches. However, we found that the k-space network is more adept at removing high-frequency oscillatory artifacts, complementing the benefits of including the k-space branch in A-LIKNet. Adding the low-rank subnetwork (A-LINet) further enhances the reconstruction performance of the image network at higher accelerations. By retaining the main low-rank components, noise can be effectively suppressed. Additionally, we observed that networks with a low-rank module learn more accurate signal intensity thanks to the noise suppression.

Furthermore, we observed the learned attention maps in the attention block before the UNet output layer in the image subnetwork (Fig.~\ref{fig:attention_visualization} in the supplementary material). The weights for each frame are not consistent, as they change with subjects, iteration numbers, and acceleration factors. However, we found that in the first two iterations, diastolic frames tend to have higher weights. We speculate that this is because, when the noise level is relatively high in the early learning stage, frames with smaller changes in diastolic phases are more conducive to structural learning. In the last two iterations, the first and last frames receive lower attention, while frames in the middle have similar importance. This could be due to the lack of adjacent frames at the beginning and end of the cardiac cycle, resulting in less usable temporal information. It is conceivable to mirror the cyclic frames, but it necessitates a larger input tensor and an augmented memory footprint. For the limited temporal receptive field of view, we would not expect a substantial difference.  

\subsection{Limitations and outlook}
While the proposed A-LIKNet demonstrates outstanding performance in cardiac Cine MR imaging reconstruction, it still exhibits certain limitations. One of the drawbacks is the parameter complexity of A-LIKNet, given that each iteration involves multiple subnetworks, resulting in increased training and reconstruction times, as summarized in Tab.~\ref{tab:summary_comparison}. Restricted by the memory limitation, the current number of iterations is limited. Future research will explore methods to enhance each component's computational efficiency while ensuring that the network performance remains uncompromised. With an efficient network, more unrolls are possible, which could provide a better performance. Additionally, we observed that the reconstruction performance of A-LIKNet for slices during the cardiac systolic phase slightly decreases. In the future, we plan to introduce motion fields for compensation. Furthermore, the current work has not yet been rolled out in the clinic and has only been validated on prospectively acquired $8\times$ accelerated OCMR data. Therefore, we plan to implement our work on the scanner to test A-LIKNet on prospectively undersampled data with higher acceleration factors.

\section{Conclusion}
This paper presented a novel deep learning-based method for dynamic MRI reconstruction named A-LIKNet. The parallel-branch architecture of A-LIKNet enables independent learning in both k-space and image domains. Utilizing multi-domain information and enabling information exchange through information sharing layers contribute to exploring and potentially converging towards global optima, as suggested by our findings. Additionally, leveraging the low-rank characteristics of dynamic imaging aids in noise suppression during the reconstruction. Furthermore, the introduction of attention mechanisms grants the network flexibility to select more valuable features over MR coils and time frames. Comparative experiments demonstrated that the proposed A-LIKNet significantly outperforms existing deep learning-based approaches for dynamic image reconstruction. Ablation studies verified the contributions of each submodule in improving network performance. Our network can reconstruct up to $24\times$ retrospectively undersampled cardiac Cine MRI with high image quality, which in turn would correspond in a prospective setting to a total scan time of 5 to 12 seconds. We found similar performance between retrospective and prospective reconstructions for $8\times$ accelerations, showcasing the potential of the proposed A-LIKNet to enable single breath-hold acquisition in the future.

\section*{Acknowledgments}
The work was supported by the Deutsche Forschungsgemeinschaft (DFG, German Research Foundation) under Germany’s Excellence Strategy – EXC 2064/1 – Project number 390727645.

\bibliographystyle{unsrt}  
\bibliography{references}

\pagebreak
\onecolumn
\begin{center}
\textbf{\large Supplemental Materials}
\end{center}
\setcounter{equation}{0}
\setcounter{figure}{0}
\setcounter{table}{0}
\setcounter{page}{1}
\makeatletter
\renewcommand{\theequation}{S\arabic{equation}}
\renewcommand{\thefigure}{S\arabic{figure}}

\begin{figure}[h!]
\centering
\includegraphics[width=0.9\textwidth]{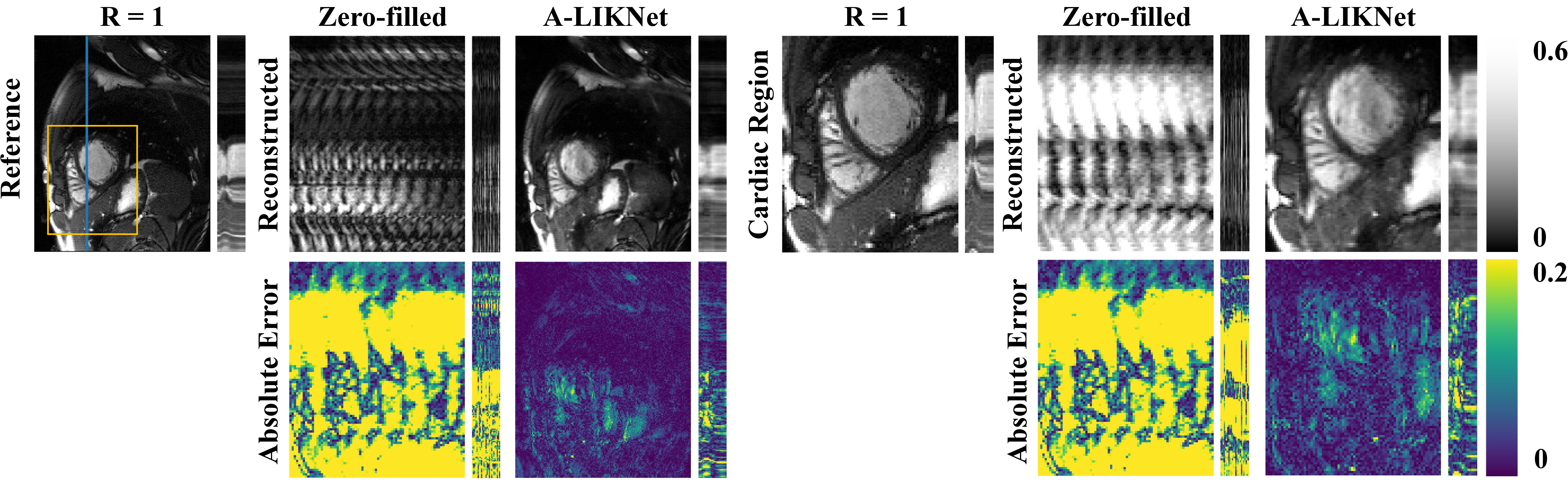}
\caption{Reconstructions in spatial (x-y) and spatial-temporal (y-t) plane of the proposed A-LIKNet for a retrospectively R=30 undersampled (VISTA) patient with transposition of the great vessels. The dynamic performance in the y-t plane corresponds to the blue line in the reference x-y plane image. Both the entire field of view (left) and the enlarged views of the cardiac region (right, yellow box region) are shown alongside the corresponding 5-times scaled absolute error maps.}
\label{fig:recon_30x}
\end{figure}

\begin{figure}[h!]
\includegraphics[width=\textwidth]{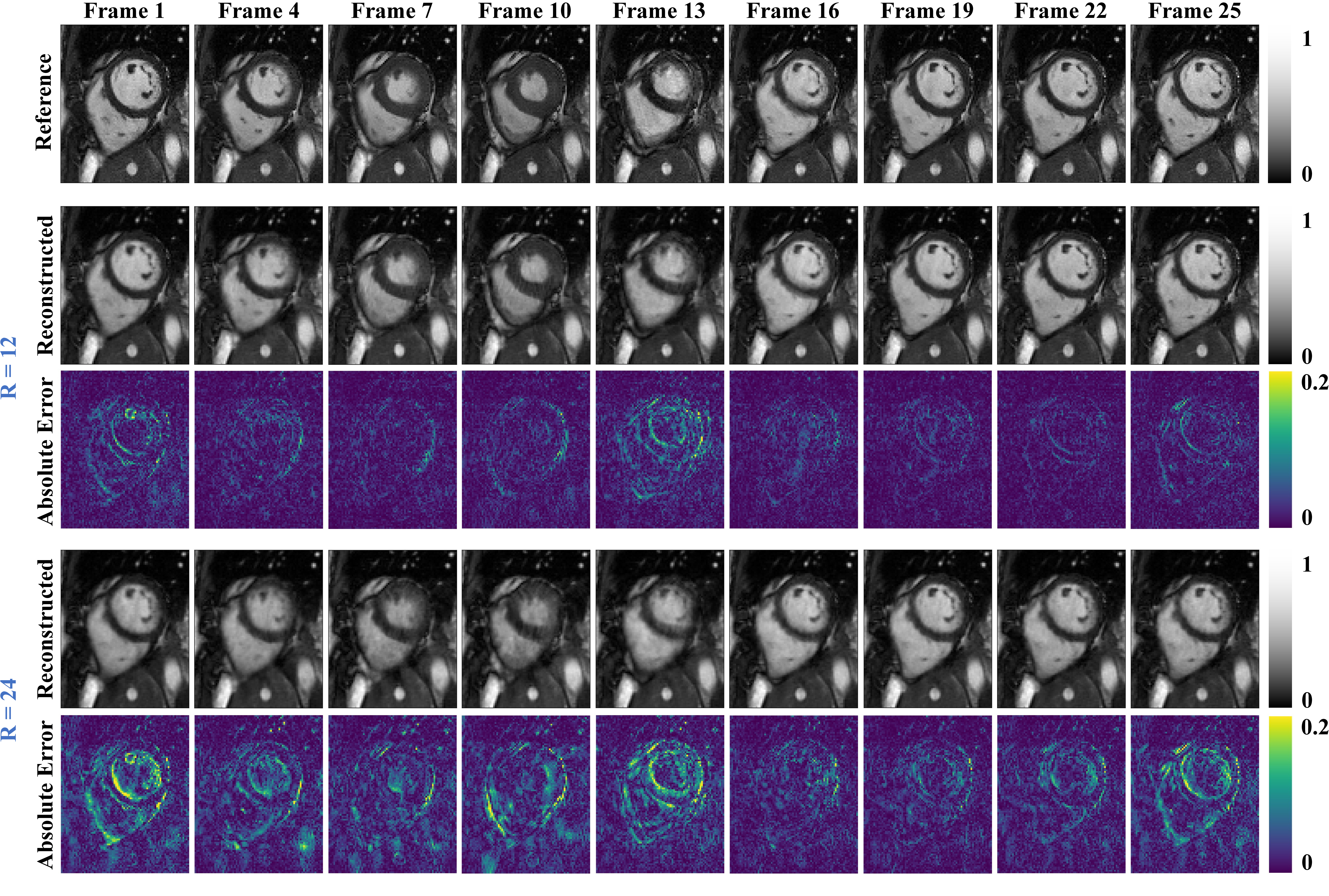}
\caption{Enlarged cardiac region of Fig.~\ref{fig:recon_results_cycle}. Reconstructions in spatial (x-y) plane of the proposed A-LIKNet for a retrospectively undersampled (VISTA) healthy subject. Results for every 3rd frame over one cardiac cycle are shown in each column. Both R=12 and R=24 reconstructions are shown alongside the corresponding absolute error maps.}
\label{fig:Recon_result_cycle_cardiac}
\end{figure}

\begin{figure}
\includegraphics[width=\textwidth]{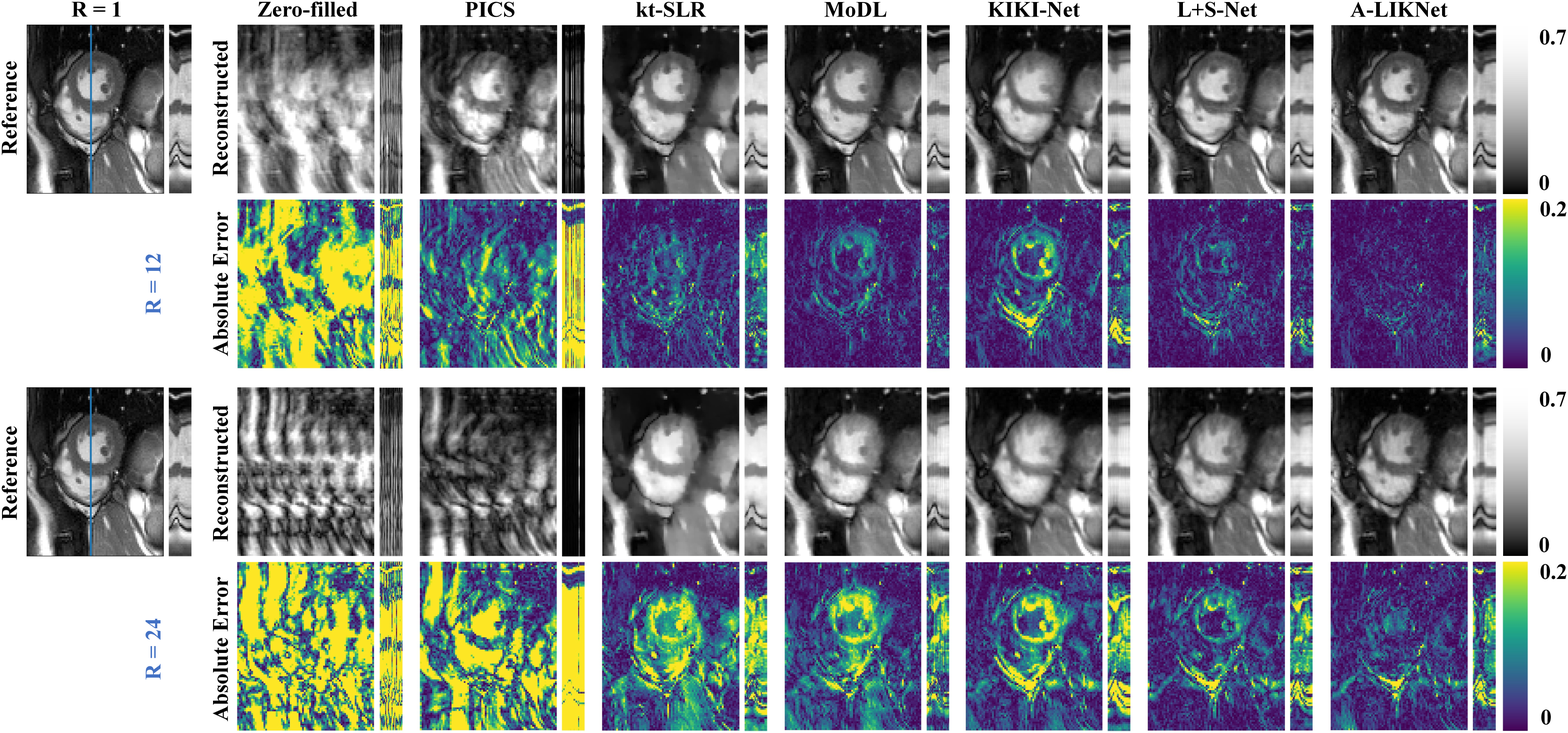}
\caption{Enlarged cardiac region of Fig.~\ref{fig:compare_results}. Reconstructions in spatial (x-y) and spatial-temporal (y-t) plane of the proposed A-LIKNet in comparison to zero-filled, PICS, kt-SLR, MoDL, KIKI-Net, and L+S-Net for a patient with active myocarditis who was retrospectively undersampled with VISTA sampling. The dynamic performance in the y-t plane corresponds to the blue line in the reference x-y plane image. Both R=12 (top) and R=24 (bottom) reconstructions are shown alongside the corresponding absolute error maps.}
\label{fig:compare_results_cardiac}
\end{figure}

\begin{figure}
\includegraphics[width=\linewidth]{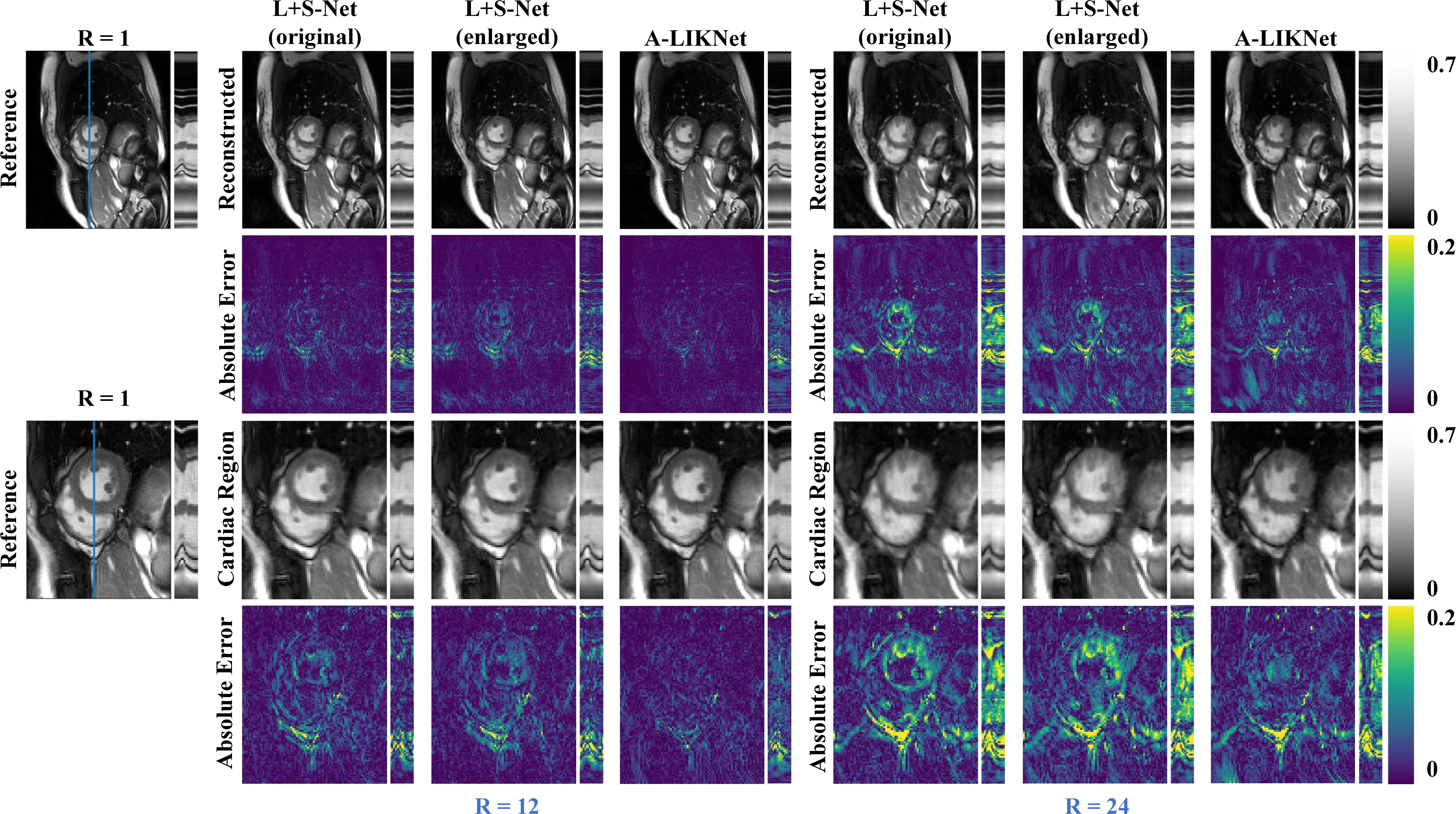}
\caption{Reconstructions in spatial (x-y) and spatial-temporal (y-t) plane of the proposed A-LIKNet in comparison to original L+S-Net and enlarged L+S-Net for a patient with active myocarditis who was retrospectively undersampled with VISTA sampling. The dynamic performance in the y-t plane corresponds to the blue line in the reference x-y plane image. Both R=12 (left) and R=24 (right) reconstructions are shown alongside the corresponding absolute error maps.}
\label{fig:LSNet_enlarged}
\end{figure}

\begin{figure}
\centering
\includegraphics[width=0.8\textwidth]{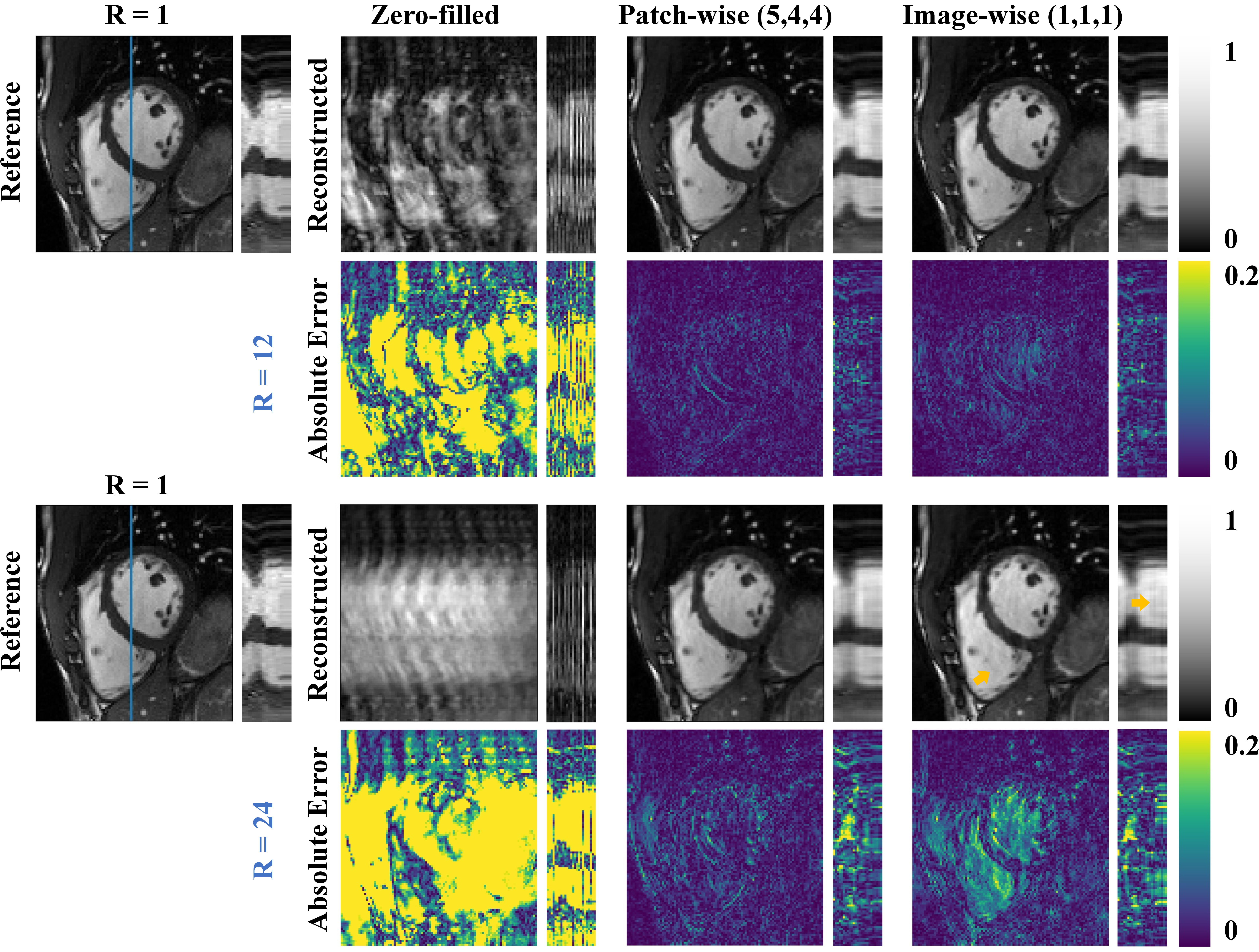}
\caption{Ablations on patch/image input of low-rank subnetwork: Reconstructions in spatial (x-y) and spatial-temporal (y-t) plane of the proposed A-LIKNet with patch-/image-wise low-rank subnetwork for a healthy subject who was retrospectively undersampled with VISTA sampling. The dynamic performance in the y-t plane corresponds to the blue line in the reference x-y plane image. Both R=12 (top) and R=24 (bottom) results are shown alongside the corresponding absolute error maps.}
\label{fig:ablation_patch}
\end{figure}

\begin{figure}
\includegraphics[width=\textwidth]{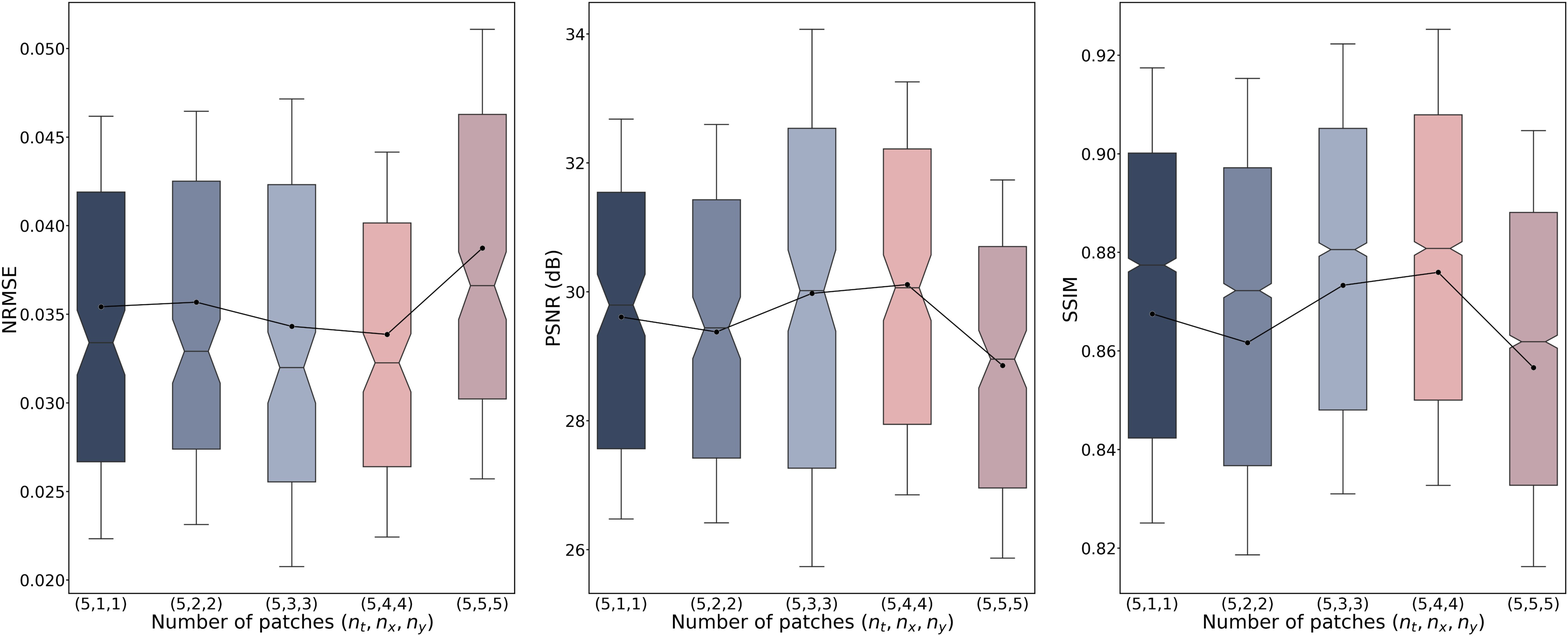}
\caption{Quantitative analysis in terms of NRMSE, PSNR, and SSIM between different spatio-temporal patch divisions in the low-rank subnetwork in A-LIKNet. Results are calculated for all subjects in the test dataset at R=24 and depicted in box plots (horizontal line: median, box: 25\% and 75\% percentile, whiskers: 0.3 $\cdot$ interquartile range). The line connects the mean values of each group of data.}
\label{fig:patch_quant}
\end{figure}

\begin{figure}
\centering
\includegraphics[width=0.7\textwidth]{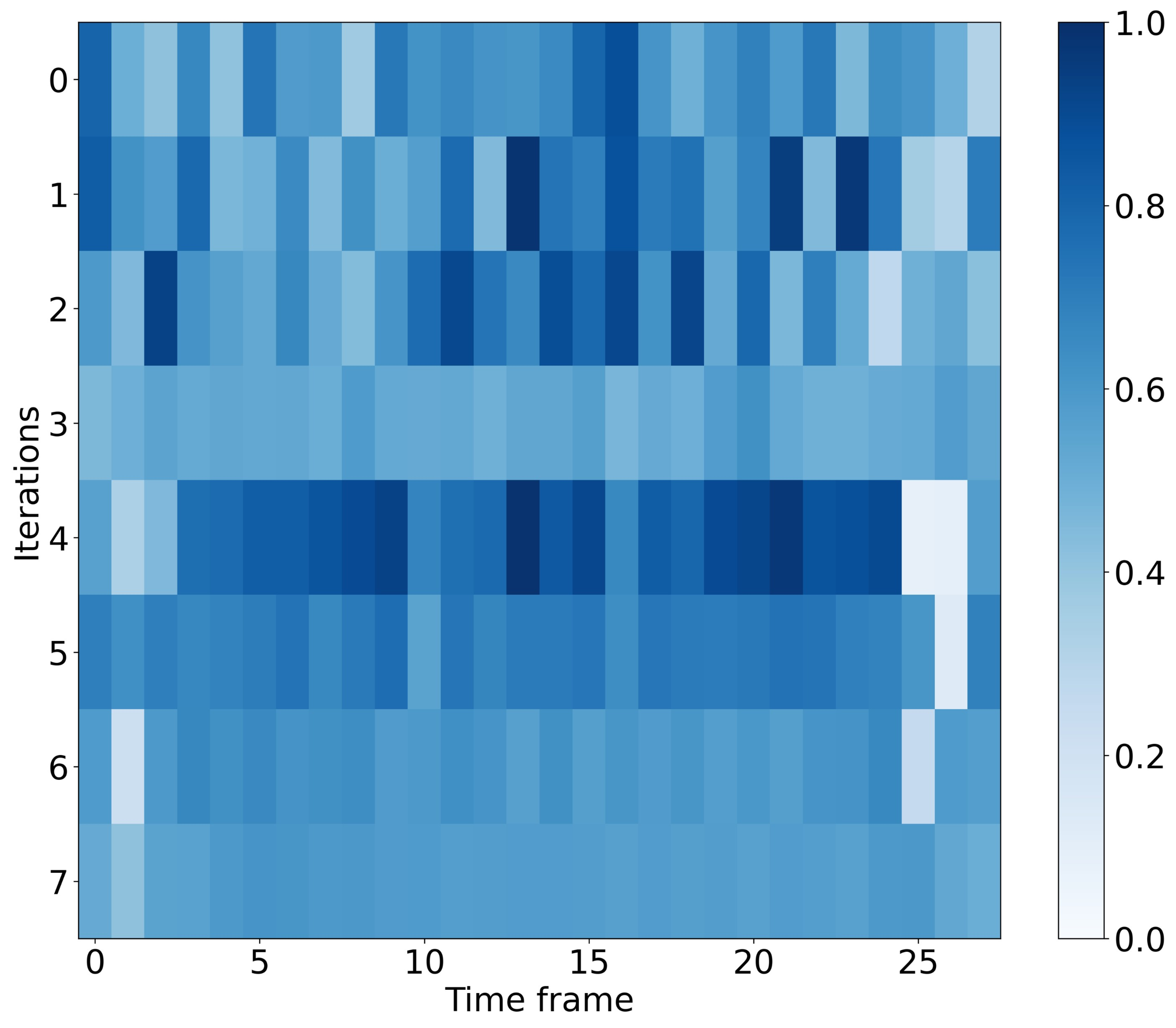}
\caption{Learned attention map in the attention block before the UNet output layer in the image subnetwork. Each row depicts all time frames (after padding) in each iteration, ranging from systolic (frame 0 to approx. 10) to diastolic phase (frame 11 to 27). We average the real and imaginary part of attention for each time frame. We observe that in the first two iterations, frames during the diastolic phase (after 10$^{th}$ time frame) tend to have higher weights. We hypothesize that in the early learning stage when the noise level is relatively high, frames with smaller changes in diastolic phases are more conducive to structural learning. In the last two iterations, the first and last frames receive lower attention, while frames in the middle have similar importance. This could be due to the lack of adjacent frames at the beginning and end of the cardiac cycle, resulting in less usable temporal information.} 
\label{fig:attention_visualization}
\end{figure}

\end{document}